\documentclass[a4paper,12pt]{article}
\usepackage{amssymb,amsmath,bm}
\usepackage{graphicx}
\usepackage{enumerate}

\title{On Gale's Contribution in Revealed Preference Theory}
\author{Yuhki Hosoya\thanks{TEL: +81-90-5525-5142, E-mail: ukki(at)gs.econ.keio.ac.jp}\\ Faculty of Economics, Chuo University\thanks{742-1, Higashinakano, Hachioji-shi, Tokyo, 192-0393, Japan.}}
\date{}
\begin{document}
\maketitle

\begin{abstract}
We investigate Gale's important paper published in 1960. This paper contains an example of a candidate of the demand function that satisfies the weak axiom of revealed preference and that is doubtful that it is a demand function of some weak order. We examine this paper and first scrutinize what Gale proved. Then we identify a gap in Gale's proof and show that he failed to show that this candidate of the demand function is not a demand function. Next, we present three complete proofs of Gale's claim. First, we construct a proof that was constructible in 1960 by a fact that Gale himself demonstrated. Second, we construct a modern and simple proof using Shephard's lemma. Third, we construct a proof that follows the direction that Gale originally conceived. Our conclusion is as follows: although, in 1960, Gale was not able to prove that the candidate of the demand function that he constructed is not a demand function, he substantially proved it, and therefore it is fair to say that the credit for finding a candidate of the demand function that satisfies the weak axiom but is not a demand function is attributed to Gale.

\vspace{12pt}

\vspace{12pt}

\vspace{12pt}
\end{abstract}

\section{Introduction}
Since the weak axiom of revealed preference was created by Samuelson (1938), there had been an active debate in consumer theory regarding what type of choice behavior can be expressed by utility maximizing behavior. When a candidate of the demand function is given, what are the conditions for it to behave as if it were a consequence of some weak order maximization problem? In the 1950s, several researchers argued that the weak axiom of revealed preference is a necessary and sufficient condition for this problem. Although Houthakker (1950) presented the strong axiom of revealed preference and argued that it was the answer to the above question, Houthakker himself does not present any example of a function that obeys the weak axiom but violates the strong axiom, and some economists seemed to believe that the weak axiom of revealed preference is sufficient.

Gale (1960) is a landmark paper that offered an answer to this controversy. He presented a candidate of the demand function and showed that, although it satisfies the weak axiom of revealed preference, it is not a consequence of any weak order maximizing behavior. This made it clear that the weak axiom of revealed preference is not sufficient to answer the question in the previous paragraph. Eventually, Richter (1966) rigorously showed that Houthakker's claim is correct, and that settled this issue.

However, some economists argue that this thesis presented by Gale is insufficient. According to them, Gale only showed that the candidate of the demand function he constructed satisfies the weak axiom of revealed preference, but he did not rigorously show that it does not correspond to any weak order. The primary objective of the present paper is to investigate this claim; that is, to investigate what Gale's contribution is and whether the example Gale constructed is really not a demand function.

We first scrutinize Gale's constructed candidate of the demand function and his discussion on this function. Gale showed that this candidate satisfies the weak axiom of revealed preference and the corresponding inverse demand function does not satisfy Jacobi's integrability condition. However, in Gale's paper, there is no sufficient proof for this function to be not able to be represented by the weak order maximizing behavior. Thus, the assertion in the previous paragraph is correct.

Next, we present three complete proofs of Gale's claim. The first is a proof using another fact that Gale himself presented in 1960. In addition to the argument above, Gale showed that the constructed candidate of the demand function does not satisfy the strong axiom of revealed preference. As mentioned above, the strong axiom of revealed preference is a necessary and sufficient condition for a function to be a demand function that corresponds to a weak order. This was proved by Richter in 1966; thus, this was not yet proved in 1960. So can Gale's claim be proved without using Richter's result but only from what was known in 1960? We answer this question affirmatively. From this point of view, we think that it is fair to say that while the assertions in the two previous paragraphs are indeed correct, the proof of Gale's claim can be easily reconstructed from what Gale himself showed, and thus it is Gale who showed that the weak axiom of revealed preference is insufficient for rationalizability.

Second, we provide a modern and sophisticated proof of Gale's claim. The above proof is correct, but Gale's proof that the function does not satisfy the strong axiom of revealed preference is too genius to be easily conceived. In this section, we provide a more mechanical, easy, and simple proof using Shephard's lemma. Perhaps, this proof is the most standard one at the present time.

Third, we prove the above result using Jacobi's integrability condition as addressed by Gale. Although both of the above proofs are simple and rigorous, they are fundamentally different from the proof scheme that Gale himself would have had in mind. Therefore, we start from the inverse demand function, construct a method for calculating a binary relation, and prove that Jacobi's integrability condition is equivalent to the transitivity of this binary relation. Because Gale's function can be seen as a consequence of the maximizing behavior of this binary relation, the fact that this binary relation is not weakly ordered means that this function cannot be rationalized. Although this proof is heavily complicated, this is probably the closest to the proof that Gale had in mind.

We prove our main claims in the main text, but we have chosen to place some of the proof in the appendix. In particular, the result of Richter (1966) (Theorem 1), Shephard's lemma (Theorem 3), and the method for constructing a binary relation from an inverse demand function (Theorem 4) are all results that are too heavy to read the proofs in situ, and thus we place the proofs of these results in the appendix.

This paper is organized as follows. In Section 2, we introduce many symbols and terms that are necessary for this paper. In Section 3, we scrutinize Gale's paper and the result that Gale showed. Section 4 contains three proofs of Gale's claim listed above. Section 5 is the conclusion.

\section{Preliminaries}

In this section, we provide basic knowledge of consumer theory. First, we define $\mathbb{R}^M_+=\{x\in \mathbb{R}^M|x_i\ge 0\mbox{ for all }i\}$ and $\mathbb{R}^M_{++}=\{x\in \mathbb{R}^M|x_i>0\mbox{ for all }i\}$. The former is called the {\bf nonnegative orthant} of $\mathbb{R}^M$, and the latter is called the {\bf positive orthant} of $\mathbb{R}^M$, respectively. If $M=1$, then this symbol is abbreviated, and these sets are simply written as $\mathbb{R}_+$ and $\mathbb{R}_{++}$. For $x,y\in\mathbb{R}^M$, we write $x\ge y$ if and only if $x-y\in \mathbb{R}^M_+$, and $x\gg y$ if and only if $x-y\in\mathbb{R}^M_{++}$.

In this paper, $\Omega$ denotes the consumption space, and we assume that $\Omega$ is a subset of $\mathbb{R}^n$. An element $x\in \Omega$ is called a {\bf consumption plan}, and $x_i$ denotes the amount of consumption for the $i$-th commodity.

Choose any binary relation $R$ on $\Omega$.\footnote{That is, $R\subset \Omega^2$.} We write $xRy$ instead of $(x,y)\in R$. We say that $R$ is\footnote{Note that, reflexivity, completeness, transitivity, symmetry, asymmetry, and antisymmetry can be defined even when $R$ is a binary relation for some abstract set $X$.}
\begin{itemize}
\item {\bf reflexive} if $xRx$ for all $x\in \Omega$,

\item {\bf complete} if for all $x,y\in \Omega$, either $xRy$ or $yRx$,

\item {\bf transitive} if $xRy$ and $yRz$ imply $xRz$,

\item {\bf p-transitive} if $\dim(\mbox{span}\{x,y,z\})\le 2$, $xRy$ and $yRz$ imply $xRz$,

\item {\bf continuous} if $R$ is closed in $\Omega^2$,

\item {\bf symmetric} if $xRy$ implies $yRx$,

\item {\bf asymmetric} if $xRy$ implies $\neg (yRx)$,

\item {\bf antisymmetric} if $xRy$ and $yRx$ imply $x=y$,

\item {\bf monotone} if $x\gg y$ implies $xRy$ and $\neg (yRx)$,

\item {\bf convex} if $xRy$, $yRx$ and $t\in [0,1]$ imply $[(1-t)x+ty]Rx$, and

\item {\bf strictly convex} if $xRy$, $yRx$, $x\neq y$, and $t\in ]0,1[$ imply $[(1-t)x+ty]Rx$ and $\neg (xR[(1-t)x+ty])$.
\end{itemize}
We call a complete and transitive binary relation a {\bf weak order}, a transitive and asymmetric binary relation a {\bf strong order}, and a reflexive, transitive, and symmetric binary relation an {\bf equivalence relation}. Suppose that $\succsim$ is a binary relation, and define
\[x\succ y\Leftrightarrow x\succsim y\mbox{ and }y\not\succsim x,\]
\[x\sim y\Leftrightarrow x\succsim y\mbox{ and }y\succsim x.\]
It is known that if $\succsim$ is a weak order, then $\succ$ is a strong order and $\sim$ is an equivalence relation.\footnote{Only transitivity of $\succ$ is not trivial, and thus we show this. Suppose not. Then, there exist $x,y,z\in \Omega$ such that $x\succ y$ and $y\succ z$ but $x\not\succ z$. Because $x\succ y$ and $y\succ z$, $x\succsim y$ and $y\succsim z$, and thus $x\succsim z$. Hence, we must have $z\succsim x$. By the transitivity of $\succsim$, we have that $z\succsim y$, which contradicts $y\succ z$.}

Choose any binary relation $\succsim$ on $\Omega$. A function $u:\Omega\to \mathbb{R}$ is said to {\bf represent} $\succsim$, or to be a {\bf utility function} of $\succsim$, if
\[x\succsim y\Leftrightarrow u(x)\ge u(y).\]
It is easy to show that if $\succsim$ is represented by a function, then it is a weak order. Debreu (1954) showed that if $\succsim$ is a continuous weak order and $\Omega$ is connected, then there exists a continuous function that represents $\succsim$.

Choose $p\in \mathbb{R}^n_{++}$ and $m>0$. Define
\[\Delta(p,m)=\{x\in \Omega|p\cdot x\le m\},\]
and for a binary relation $\succsim$ defined on $\Omega$, define
\[f^{\succsim}(p,m)=\{x\in \Delta(p,m)|y\not\succ x\mbox{ for all }y\in \Delta(p,m)\}.\]
We call this multi-valued function $f^{\succsim}$ a {\bf semi-demand function} corresponds to $\succsim$, and if $\succsim$ is a weak order, then we call $f^{\succsim}$ a {\bf demand function} corresponds to $\succsim$. If $\succsim$ is represented by $u$, then $f^{\succsim}$ is also written as $f^u$.

Note that, for given $(p,m)\in \mathbb{R}^n_{++}\times \mathbb{R}_{++}$, $f^u(p,m)$ is the set of solutions to the following maximization problem:
\begin{align}
\max~~~~~&~u(x)\nonumber \\
\mbox{subject to. }&~x\in \Omega,\label{UMP}\\
&~p\cdot x\le m.\nonumber
\end{align}
We call this problem (\ref{UMP}) the {\bf utility maximization problem} for the utility function $u$. Therefore, the demand function $f^u$ is explained as the (possibly multi-valued) solution function to the utility maximization problem (\ref{UMP}).

Suppose that $f:P\to \Omega$, where $P$ is a nonempty cone that is included in the following set:
\[\{(p,m)\in \mathbb{R}^n_{++}\times \mathbb{R}_{++}|\Delta(p,m)\neq \emptyset\}.\]
We call this function $f$ a {\bf candidate of demand} (CoD) if $f(p,m)\in \Delta(p,m)$ for all $(p,m)\in P$. If $f=f^{\succsim}$, then we say that $f$ corresponds to $\succsim$ or $\succsim$ corresponds to $f$. Moreover, if $\succsim$ is represented by $u$, then we say that $f$ corresponds to $u$ or $u$ corresponds to $f$. We call a CoD $f$ a {\bf semi-demand function} if it is a semi-demand function corresponds to some binary relation $\succsim$, and a {\bf demand function} if it is a demand function corresponds to some weak order $\succsim$.

For a CoD $f$, let $R(f)$ be the range of $f$: that is,
\[R(f)=\{x\in \Omega|x=f(p,m)\mbox{ for some }(p,m)\in P\}.\]

We say that a CoD $f$ is {\bf homogeneous of degree zero} if
\[f(ap,am)=f(p,m)\]
for any $a>0$. Note that, because $P$ is a cone, the above definition has no problem. Next, we say that a CoD $f$ satisfies {\bf Walras' law} if
\[p\cdot f(p,m)=m\]
for any $(p,m)\in P$. Note that, if $f=f^{\succsim}$ for some weak order $\succsim$, then $f$ is automatically homogeneous of degree zero. Moreover, if $\succsim$ is monotone, then $f$ automatically satisfies Walras' law.

For a CoD $f$, define a binary relation $\succ_r$ on $\Omega$ such that
\[x\succ_ry\Leftrightarrow \exists (p,m)\in P\mbox{ s.t. }f(p,m)=x,\ y\in \Delta(p,m)\setminus\{x\}.\]
We call this relation $\succ_r$ the {\bf direct revealed preference relation}, and we say that $f$ satisfies the {\bf weak axiom of revealed preference} (or simply, the {\bf weak axiom}) if $\succ_r$ is asymmetric.

Choose any binary relation $R$ on $\Omega$, and define
\[R^*=\cap\{\bar{R}\subset \Omega^2|R\subset \bar{R}\mbox{ and }\bar{R}\mbox{ is transitive}\}.\]
This binary relation $R^*$ is called the {\bf transitive closure} of $R$. Because the intersection of any family of transitive binary relations is also transitive, we have that $R^*$ is transitive. In the appendix, we show that $xR^*y$ if and only if there exists a finite sequence $x_1,...,x_k$ such that $x_1=x,x_k=y$ and $x_iRx_{i+1}$ for all $i\in \{1,...,k-1\}$.

Specifically, let $\succ_{ir}$ be the transitive closure of $\succ_r$. We call this relation $\succ_{ir}$ the {\bf indirect revealed preference relation}, and we say that $f$ satisfies the {\bf strong axiom of revealed preference} (or simply, the {\bf strong axiom}) if $\succ_{ir}$ is asymmetric. Clearly, the strong axiom implies the weak axiom.

Finally, suppose that $f:P\to \Omega$ is a CoD that is differentiable at $(p,m)$. Define
\[s_{ij}(p,m)=\frac{\partial f_i}{\partial p_j}(p,m)+\frac{\partial f_i}{\partial m}(p,m)f_j(p,m).\]
The $n\times n$ matrix $S_f(p,m)$ whose $(i,j)$-th component is $s_{ij}(p,m)$ is called the {\bf Slutsky matrix}. We sometimes call the matrix-valued function $S_f$ itself the Slutsky matrix. This matrix has the following alternative notation:
\[S_f(p,m)=D_pf(p,m)+D_mf(p,m)f^T(p,m),\]
where $D_p$ and $D_m$ denote the partial differential operators, and $f^T$ denotes the transpose of $f$.\footnote{Throughout this paper, $A^T$ denotes the transpose of the matrix $A$.}

\section{Gale's Example}
Gale (1960) presented a CoD that satisfies the weak axiom of revealed preference but is doubtful to be a demand function. In this section, we introduce Gale's example. Throughout this section, we consider $\Omega=\mathbb{R}^n_+$ and $P=\mathbb{R}^n_{++}\times\mathbb{R}_{++}$, and usually assume that $n=3$.

Let
\[A=\begin{pmatrix}
-3 & 4 & 0\\
0 & -3 & 4\\
4 & 0 & -3
\end{pmatrix},\]
and define
\[h_A(p)=\frac{1}{p^TAp}Ap\]
on the set
\[C=\{p\in\mathbb{R}^3_{++}|Ap\ge 0\}.\]
Note that, if $p\in C$, then $Ap\neq 0$ and $p^TAp>0$, and thus $h_A(p)$ is well-defined. Indeed, if $p\in \mathbb{R}^3_{++}$ and $Ap\le 0$, then
\[4p_2\le 3p_1,\ 4p_3\le 3p_2,\ 4p_1\le 3p_3,\]
and thus, $64p_1\le 27p_1$, which is a contradiction. Hence, if $p\in C$, then $Ap\ge 0$ and $Ap\neq 0$, which implies that $p^TAp>0$.

Now, choose any $p\in \mathbb{R}^3_{++}$, and define $\bar{p}$ as follows.
\begin{enumerate}[I)]
\item If $p\in C$, then we define $\bar{p}=p$.

\item Suppose that for some $(i,j,k)\in \{(1,2,3),(2,3,1),(3,1,2)\}$, $-3p_i+4p_j\le 0$, and $-3p_j+4p_k\le 0$. By our previous argument, we have that $-3p_k+4p_i>0$. In this case, define $\bar{p}_i=\frac{16}{9}p_k$, $\bar{p}_j=\frac{4}{3}p_k$, and $\bar{p}_k=p_k$.

\item Suppose that for some $(i,j,k)\in \{(1,2,3),(2,3,1),(3,1,2)\}$, $-3p_i+4p_j\le 0$, $-3p_j+4p_k\ge 0$, and $-3p_k+4p_i\ge 0$. We separate this case into two subcases.
\begin{enumerate}[i)]
\item If $16p_j-9p_k\ge 0$, then define $\bar{p}_i=\frac{4}{3}p_j$, $\bar{p}_j=p_j$, and $\bar{p}_k=p_k$.

\item If $16p_j-9p_k\le 0$, then define $\bar{p}_i=\frac{4}{3}p_j$, $\bar{p}_j=p_j$, and $\bar{p}_k=\frac{16}{9}p_j$.
\end{enumerate}
\end{enumerate}
In the appendix, we check that $\bar{p}$ is well-defined and the mapping $p\mapsto \bar{p}$ is continuous. Note that, in any case, $\bar{p}\le p$ and $\bar{p}\in C$. Moreover, $\bar{p}$ is homogeneous of degree one with respect to $p$: that is, $\overline{ap}=a\bar{p}$ for all $a>0$. Define
\begin{equation}\label{GALE}
f^G(p,m)=h_A(\bar{p})m.
\end{equation}
We call this CoD $f^G$ {\bf Gale's example}.

It is easy to show that Gale's example $f^G$ is a continuous CoD that is homogeneous of degree zero and satisfies Walras' law. We show that $f^G$ satisfies the weak axiom. Suppose not. Then, there exist $x,y\in \Omega$ such that $x\succ_ry$ and $y\succ_rx$. Therefore, $x\neq y$, and there exist $(p,m),(q,w)\in P$ such that $x=f^G(p,m),\ p\cdot y\le m$ and $y=f^G(q,w),\ q\cdot x\le w$. Because Gale's example is homogeneous of degree zero, replacing $(p,m),(q,w)$ with $\frac{1}{m}(p,m),\frac{1}{w}(q,w)$, we can assume without loss of generality that $m=w=1$. Moreover, because $\bar{p}\le p$ and $\bar{q}\le q$, we can assume without loss of generality that $p=\bar{p},\ q=\bar{q}$. Then,
\[q^TAp\le p^TAp,\ p^TAq\le q^TAq.\]
Define
\[\lambda=\frac{p^TAp}{q^TAp}.\]
By definition, $\lambda\ge 1$. Define $r=\lambda q$. Then, $(r-p)^TAp=0$, and
\[p^TAr=\lambda p^TAq\le \lambda^2p^TAq\le \lambda^2q^TAq=r^TAr,\]
which implies that
\[0\le (r-p)^TAr=(r-p)^TA(r-p).\]
Let $z=r-p$. By the above arguments,
\[z^TAz\ge 0,\ z^TAp=0.\]
If $z\neq 0$, then for some $(i,j,k)\in \{(1,2,3), (2,3,1), (3,1,2)\}$, $z_i\ge 0, z_k\le 0$ and $z_i-z_k>0$. If $z_j\ge 0$, then
\begin{align*}
0\le z^TAz=&~-3z_i^2-3z_j^2-3z_k^2+4z_iz_j+4z_jz_k+4z_kz_i\\
=&~-(3z_i^2-4z_iz_j+3z_j^2)-3z_k^2+4z_k(z_i+z_j)<0,
\end{align*}
which is a contradiction. If $z_j\le 0$, then
\[0\le z^TAz=-(3z_j^2-4z_jz_k+3z_k^2)-3z_i^2+4z_i(z_j+z_k)<0,\]
which is a contradiction. Therefore, we have that $z=0$. This implies that $q$ is proportional to $p$, and thus $x=y$, which is a contradiction. Hence, $f^G$ satisfies the weak axiom.

Gale claimed that $f^G$ is not a demand function. The reason he explained is as follows. First, suppose that $f:P\to \Omega$ is a CoD, and $g:\mathbb{R}^n_{++}\to \mathbb{R}^n_{++}$ satisfies
\[x=f(g(x),g(x)\cdot x)\]
for every $x\in \mathbb{R}^n_{++}$. We call this function $g$ an {\bf inverse demand function} of $f$. Gale claimed that if $f$ is a demand function, then $g(x)$ must satisfy the following {\bf Jacobi's integrability condition}:\footnote{Here, we abbreviate variables to avoid the expression becoming too long. Note that, we say that $g(x)$ satisfies (\ref{JACOBI}) if and only if $g$ satisfies (\ref{JACOBI}) for all $x$ and $i,j,k\in \{1,...,n\}$, and $g(x)$ violates (\ref{JACOBI}) if there exist $x$ and $i,j,k\in \{1,...,n\}$ such that (\ref{JACOBI}) is not satisfied.}

\begin{equation}\label{JACOBI}
g_i\left(\frac{\partial g_j}{\partial x_k}-\frac{\partial g_k}{\partial x_j}\right)+g_j\left(\frac{\partial g_k}{\partial x_i}-\frac{\partial g_i}{\partial x_k}\right)+g_k\left(\frac{\partial g_i}{\partial x_j}-\frac{\partial g_j}{\partial x_i}\right)=0.
\end{equation}

We define
\[B\equiv A^{-1}=\frac{1}{37}\begin{pmatrix}
9 & 12 & 16\\
16 & 9 & 12\\
12 & 16 & 9
\end{pmatrix}.\]
Then, $Bx\ge 0$ for all $x\ge 0$, and $Bx\gg 0,\ x^TBx>0$ if $x\neq 0$. Define
\begin{equation}\label{GALE2}
g(x)=\frac{1}{x^TBx}Bx.
\end{equation}
Then, for all $x\in \mathbb{R}^3_{++}$,
\[h_A(g(x))=\frac{(x^TBx)^2}{x^TB^TABx}\frac{1}{x^TBx}ABx=x,\]
which implies that $g(x)\in C$, and thus 
\[f^G(g(x),g(x)\cdot x)=x.\]
However, this $g(x)$ violates (\ref{JACOBI}), and thus, if Gale's claim is correct, then $f^G$ is not a demand function.\footnote{We verify the fact that $g(x)$ violates (\ref{JACOBI}) in the appendix. See Subsection A.5.}

We point out the problem included in Gale's explanation. First, Gale's explanation is deeply related to the following theorem.

\vspace{12pt}
\noindent
{\bf Frobenius' Theorem}. Suppose that $U\subset \mathbb{R}^n$ and $g:U\to \mathbb{R}^n\setminus\{0\}$ is $C^1$. Then, the following two statements are equivalent.
\begin{enumerate}[i.]
\item $g(x)$ satisfies (\ref{JACOBI}).

\item For every $x\in U$, there exist a neighborhood $V\subset U$ of $x$ and a pair of a $C^1$ function $u:V\to \mathbb{R}$ and a positive continuous function $\lambda:V\to \mathbb{R}$ such that
\begin{equation}\label{FROBENIUS}
\nabla u(y)=\lambda(y)g(y)
\end{equation}
for all $y\in V$.
\end{enumerate}

\vspace{12pt}
For a proof of this theorem, see Theorem 2 of Hosoya (2021a) and Theorem 10.9.4 of Dieudonne (1969).

If $f^G=f^u$ for some $C^1$ nondegenerate function $u$,\footnote{A $C^1$ function $u$ is said to be {\bf nondegenerate} if $\nabla u(x)\neq 0$ for all $x$.} then by Lagrange's multiplier rule, (\ref{FROBENIUS}) must hold, which implies that $g(x)$ satisfies (\ref{JACOBI}). As we have checked, $g(x)$ actually violates (\ref{JACOBI}). By the contrapositive of the above argument, we have that there is no $C^1$ nondegenerate function $u$ such that $f^G=f^u$. From this fact, Gale stated that $f^G$ is not a demand function.\footnote{Note that, Gale's construction to this example is not a guesswork. Actually, Gale defined $h_A(p)=\frac{1}{p^TAp}Ap$ for an arbitrary $3\times 3$ matrix $A$, and proved that $g(x)$ defined by (\ref{GALE2}) satisfies (\ref{JACOBI}) if and only if $A$ is symmetric. See his Lemma. And then, he tried to construct a matrix $A$ that is asymmetric but $f(p,m)=h_A(p)m$ satisfies the weak axiom, and obtained the above matrix $A$.}

However, this does not mean that there is no weak order $\succsim$ such that $f^G=f^{\succsim}$, because there exists a weak order that cannot be represented by a utility function.\footnote{In fact, Gale stated that ``$f^G$ is not a demand function corresponds to some utility function $u$.'' However, even this claim is not obvious if $u$ is not differentiable.} Therefore, we cannot determine whether $f^G$ is a demand function by Gale's explanation alone. With this consideration, we believe that Gale only showed that $f^G$ is not a demand function corresponding to some $C^1$ nondegenerate utility function $u$.

In the next section, we consider this problem, and treat three approaches to this problem. First, we use Gale's another contribution to show that $f^G$ is not a demand function. Second, we construct a proof that $f^G$ is not a demand function using Shephard's lemma. Third, we use Hosoya's (2013) integrability result to show that a restriction of $f^G$ into $(f^G)^{-1}(\mathbb{R}^3_{++})$ is actually not a demand function but a semi-demand function corresponds to a complete, p-transitive, and continuous preference relation, and construct an alternative proof that $f^G$ is not a demand function.

\section{Three Proofs of Gale's Claim}

\subsection{First Proof: Direct Method}
In this subsection, we prove that $f^G$ is not a demand function using a fact that Gale found. Gale considered the following four price vectors and consumption vectors.
\[p^1=(9,16,12),\ x^1=f^G(p^1,9)=(1,0,0),\]
\[p^2=(340,440,330),\ x^2=f^G(p^2,303)=(0.6,0,0.3),\]
\[p^3=(410,400,300),\ x^3=f^G(p^3,303)=(0.3,0,0.6),\]
\[p^4=(16,12,9),\ x^4=f^G(p^4,9)=(0,0,1).\]
We can easily check that
\[p^1\cdot x^2=9,\ p^2\cdot x^3=300,\ p^3\cdot x^4=300,\]
and thus
\[x^1\succ_rx^2,\ x^2\succ_rx^3,\ x^3\succ_rx^4.\]
Because $\succ_r\subset \succ_{ir}$,
\[x^1\succ_{ir}x^2,\ x^2\succ_{ir}x^3,\ x^3\succ_{ir}x^4,\]
and because $\succ_{ir}$ is transitive, we conclude that
\[(1,0,0)\succ_{ir}(0,0,1).\]
Gale stated that, by almost the same arguments, one can show that $(0,0,1)\succ_{ir}(0,1,0)$ and $(0,1,0)\succ_{ir}(1,0,0)$, which means that $\succ_{ir}$ is not asymmetric and $f^G$ violates the strong axiom. Actually, his claim is correct. Define
\[p^5=(440,330,340),\ x^5=f^G(p^5,303)=(0,0.3,0.6),\]
\[p^6=(400,300,410),\ x^6=f^G(p^6,303)=(0,0.6,0.3),\]
\[p^7=(12,9,16),\ x^7=f^G(p^7,9)=(0,1,0),\]
\[p^8=(330,340,440),\ x^8=f^G(p^8,303)=(0.3,0.6,0),\]
\[p^9=(300,410,400),\ x^9=f^G(p^9,303)=(0.6,0.3,0),\]
\[x^{10}=x^1=(1,0,0).\]
Then, we can easily check that $x^i\succ_rx^{i+1}$ for $i\in \{1,...,9\}$, and thus
\[(0,0,1)\succ_{ir}(1,0,0),\]
which implies that the strong axiom does not hold.

Gale claimed that because $f^G$ is not a demand function, it must violate the strong axiom by Houthakker's (1950) theorem, and Gale verified this result by constructing the above sequence. However, we claim that the converse relationship is more important. Suppose that $f^G=f^{\succsim}$ for a weak order $\succsim$, and $x\succ_ry$. By the definition of $f^{\succsim}$, we have that $x\succ y$. Because $\succ$ is transitive,
\[x^1\succ x^4,\ x^4\succ x^{10}.\]
However, because $x^{10}=x^1$, this contradicts the asymmetry of $\succ$. This implies that such a weak order $\succsim$ is absent, and thus $f^G$ is not a demand function. This is one approach for proving that Gale's example does not become a demand function.

In this context, Richter (1966) presented the following elegant theorem.

\vspace{12pt}
\noindent
{\bf Theorem 1}. Let $P=\{(p,m)\in \mathbb{R}^n_{++}\times \mathbb{R}_{++}|\Delta(p,m)\neq \emptyset\}$, and suppose that $f:P\to \Omega$ is a CoD. Then, $f$ is a demand function if and only if $f$ satisfies the strong axiom of revealed preference.

\vspace{12pt}
We present the proof of this theorem in the appendix.

We make several remarks on the above arguments. First, some readers might think that using Houthakker's result, Gale actually proved that $f^G$ is not a demand function. However, we believe this consideration is wrong. Houthakker's paper itself contains ambiguities characteristic of older economics papers, and it is debatable as to what his theorem is. As far as we can see, Houthakker's paper seems to calculate a so-called indifference hypersurface, and in this argument, he treated some kind of differential equation. Therefore, we believe that Houthakker's paper, like Gale's, implicitly assumed the differentiability of $u$. If this consideration is correct, then Houthakker's theorem cannot be used to show $f^G$ is not a demand function. However, even if our consideration is wrong, it is reasonable to assume that Gale could not show his result using Houthakker's theorem. Gale stated that ``Since the demand function in our example does not come from a preference relation it follows from Houthakker's theorem that it must fail to satisfy the Strong Axiom.'' A straightforward understanding of this statement suggests that Gale considered Houthakker's theorem to be the assertion that ``if a CoD is not a demand function, then it does not satisfy the strong axiom.'' And, as we saw above, the claim needed to complete the proof of Gale's result is the opposite of this, namely, that ``if the CoD does not satisfy the strong axiom, then it is not a demand function.''

Theorem 1 resolves this problem perfectly, because there is no vague claim. However, Richter's paper was published in 1966. Therefore, in 1960, this result was not known, and thus Gale could not apply it to $f^G$.

By the way, Gale mentioned Uzawa's (1959) result. His result is as follows.

\vspace{12pt}
\noindent
{\bf Theorem 2}. Suppose that $\Omega=\mathbb{R}^n_+$, $P=\mathbb{R}^n_{++}\times\mathbb{R}_{++}$, and a CoD $f$ satisfies the following assumptions.
\begin{enumerate}[I)]
\item $R(f)=\Omega\setminus\{0\}$.

\item $f$ satisfies Walras' law.

\item For all $p\in \mathbb{R}^n_{++}$, there exist $\varepsilon>0$ and $L>0$ such that if $\|q-p\|<\varepsilon$ and $m,w>0$, then $\|f(q,m)-f(q,w)\|\le L|m-w|$.\footnote{Uzawa originally stated this condition as follows: ``for all $(p,m)\in P$, there exist $\varepsilon>0$ and $L>0$ such that if $\|q-p\|<\varepsilon$ and $|w_i-m|<\varepsilon$ for $i\in \{1,2\}$, then $\|f(q,w_1)-f(q,w_2)\|\le L|w_1-w_2|$''. However, to the best of our understanding, Uzawa's condition is too weak to prove this theorem and there is a gap, unless our strengthened condition is used.}

\item $f$ satisfies the strong axiom of revealed preference.
\end{enumerate}
Define
\[\succsim=\{(x,y)\in \Omega^2|y\not\succ_{ir}x\}.\]
Then, $\succsim$ is a continuous weak order and $f=f^{\succsim}$.

\vspace{12pt}
We omit the proof of this theorem. Note that Gale's example $f^G$ satisfies I)-III) of this theorem. Therefore, in 1960, Gale knew that if $f^G$ satisfies the strong axiom, then $f^G$ is a demand function. However, as we noted above, the converse relationship is important for proving the absence of a weak order $\succsim$ such that $f^G=f^{\succsim}$. Therefore, this result is also not able to be used to prove that $f^G$ is not a demand function.

In the 21st century, we can easily show that Gale's example is not a demand function using Theorem 1. In 1960, proving this result was not easy. However, at least this result can be shown using Gale's sequence $x^1,...,x^{10}$, and the fact that $\succ_r\subset \succ$. We think that Gale could have constructed this logic even in 1960. In conclusion, we can say the following. First, there is a gap in Gale's proof that Gale's example is not a demand function. Second, Gale would have been able to prove this in a different manner, even in 1960. Therefore, it is fair to say that the credit for finding a CoD that satisfies the weak axiom but is not a demand function is attributed to Gale.

\subsection{Second Proof: Using Shephard's Lemma}
In the previous subsection, the sequence $x^1,...,x^{10}$ was given. The construction of such a sequence is not easy in general, and probably this could be found because Gale is a genius.\footnote{Gale himself stated that the finding of this sequence is a ``considerable labour''.} However, we and almost all readers are not geniuses, and thus finding such a sequence is an incredibly hard task.

Hence, we want to find another proof that can be constructed by usual economists. One way to find such a proof is to use Shephard's lemma and the Slutsky matrix.

To introduce the Shephard's lemma, we first consider the following minimization problem.
\begin{align}
\min~~~~~&~p\cdot y,\nonumber \\
\mbox{subject to. }&~y\in \Omega,\label{EMP}\\
&~u(y)\ge u(x).\nonumber
\end{align}
This problem (\ref{EMP}) is the dual problem for the utility maximization problem (\ref{UMP}). The problem (\ref{EMP}) is called the {\bf expenditure minimization problem}. We define the value of this problem as $E^x(p)$, and call the function $E^x$ the {\bf expenditure function}.

It is useful to consider another definition of the expenditure function. Suppose that a weak order $\succsim$ is given. Then, the expenditure function $E^x$ is defined as follows.
\begin{equation}
E^x(p)=\inf\{p\cdot y|y\succsim x\}.\label{EX}
\end{equation}
Note that, if $\succsim$ is represented by $u$, then $E^x(p)$ becomes the value of the problem (\ref{EMP}), and thus these two definitions of the function $E^x$ coincide.

The following theorem was shown in Lemma 1 of Hosoya (2020).

\vspace{12pt}
\noindent
{\bf Theorem 3}. Suppose that either $\Omega=\mathbb{R}^n_+$ or $\Omega=\mathbb{R}^n_{++}$, $P=\mathbb{R}^n_{++}\times \mathbb{R}_{++}$, and $f:P\to \Omega$ is a continuous demand function. Suppose also that $\succsim$ is a weak order such that $f=f^{\succsim}$, and define $E^x$ by (\ref{EX}). Then, $E^x:\mathbb{R}^n_{++}\to \mathbb{R}_+$ is concave and continuous. Moreover, if $f$ satisfies Walras' law and $x\in R(f)$, then the following facts hold.
\begin{enumerate}[1)]
\item $E^x(p)>0$ for all $p\in \mathbb{R}^n_{++}$.

\item If $x=f(p^*,m^*)$, then $E^x(p^*)=m^*$.

\item For every $p\in \mathbb{R}^n_{++}$, the following equality holds.
\begin{equation}\label{Shephard}
\nabla E^x(p)=f(p,E^x(p)).
\end{equation}
\end{enumerate}

\vspace{12pt}
We provide the proof of this result in the appendix. The equation (\ref{Shephard}) is famous and called {\bf Shephard's lemma}. Note that, because $f$ is continuous, we have that $E^x$ is $C^1$. If $f$ is continuously differentiable at $(p^*,m^*)$, then by 2) and 3), we can easily verify that $E^x$ is twice continuously differentiable at $p^*$, and
\[D^2E^x(p^*)=S_f(p^*,m^*),\]
where $D^2E^x(p^*)$ denotes the Hessian matrix of $E^x$ at $p^*$. By Young's theorem, $D^2E^x(p^*)$ is symmetric, and thus we conclude that $S_f(p^*,m^*)$ is also symmetric.

Consider applying this result to Gale's example $f^G$. It is easy to show that $R(f^G)$ includes $\mathbb{R}^3_{++}$, and if $(p,m)\in f^{-1}(\mathbb{R}^3_{++})$, then $f^G$ is continuously differentiable at this point. Hence, choose $p^*=(1,1,1),\ m^*=3$, and $x=(1,1,1)$. Then, $x=f^G(p^*,m^*)$. If $f^G$ is a demand function, then by the above consideration,
\[S_f(p^*,m^*)=D^2E^x(p^*)\]
is symmetric. However,
\[s_{12}(p^*,m^*)=\frac{11}{3}\neq -\frac{1}{3}=s_{21}(p^*,m^*)\]
and thus, $f^G$ is not a demand function.

This method of proof does not require any inspiration of a genius and shows that Gale's example is not a demand function in the shortest way. To construct this proof, we need only Theorem 3 and a simple calculation for $s_{ij}$.

\subsection{Third Proof: On the ``Open Cycle'' Theoretic Approach}

In this subsection, we investigate another aspect of Gale's example. In previous subsections, we proved that Gale's example is not a demand function. Consider the restriction $\tilde{f}$ of Gale's example to the inverse image of the positive orthant. Then, we can construct a complete, p-transitive, and continuous binary relation $\succsim$ such that $\tilde{f}=f^{\succsim}$. Thus, the restriction of Gale's example is in fact a semi-demand function. Moreover, our construction method for $\succsim$ has two features. First, for our construction method, $\tilde{f}$ is a demand function if and only if our $\succsim$ is transitive. Second, (\ref{JACOBI}) for an inverse demand function is equivalent to the transitivity of $\succsim$. Hence, we can conclude that Gale's example is not a demand function because $g(x)$ defined in (\ref{GALE2}) does not satisfy (\ref{JACOBI}). This explanation seems to be very close to Gale's original explanation, and thus we think that this is the form of proof that Gale originally intended to present. Furthermore, this proof is deeply related to the mysterious ``open cycle theory'' due to Pareto in 1906.

We start from recalling the notion of the {\bf inverse demand function}. Choose any function $g:\mathbb{R}^n_{++}\to \mathbb{R}^n_{++}$. This function $g$ is called an inverse demand function of a CoD $f:P\to \Omega$ if and only if $x=f(g(x),g(x)\cdot x)$ for all $x\in \mathbb{R}^n_{++}$. If $f=f^u$ for some continuous utility function $u:\Omega\to \mathbb{R}$ that is increasing, $C^1$, and nondegenerate on $\mathbb{R}^n_{++}$,\footnote{A function $u$ is said to be {\bf increasing} if $x\gg y$ implies $u(x)>u(y)$. Note that, if $f=f^u$ for some increasing function $u$, then $f$ must satisfy Walras' law.} then by Lagrange's multiplier rule, $g$ is an inverse demand function of $f$ if and only if
\[\nabla u(x)=\lambda(x)g(x)\]
for some $\lambda(x)>0$.

Define $\tilde{\Omega}=\mathbb{R}^n_{++}$. Let $g:\tilde{\Omega}\to \mathbb{R}^n_{++}$ be a given $C^1$ function. Choose any $(x,v)\in \tilde{\Omega}^2$, and consider the following ordinary differential equation (ODE):
\begin{equation}\label{IND}
\dot{y}(t)=(g(y(t))\cdot x)v-(g(y(t))\cdot v)x,\ y(0)=x.
\end{equation}
Let $y(t;x,v)$ denote the nonextendable solution for the above parametrized ODE. Define
\[w^*=(v\cdot x)v-(v\cdot v)x,\]
\[t(x,v)=\inf\{t\ge 0|y(t)\cdot w^*\ge 0\}.\]
In the appendix, we show that if $x$ is not proportional to $v$, then $t(x,v)$ is well-defined and positive, and $y(t;x,v)$ is proportional to $v$ if and only if $t=t(x,v)$. Define
\[u^g(x,v)=\frac{\|y(t(x,v);x,v)\|}{\|v\|},\]
\[\succsim^g=(u^g)^{-1}([1,+\infty[).\]
The following result is a slight modification of Theorem 1 of Hosoya (2013).

\vspace{12pt}
\noindent
{\bf Theorem 4}. Suppose that $k\ge 1$ and $g:\tilde{\Omega}\to \mathbb{R}^n_{++}$ is a $C^k$ function. Then, the following results hold.\footnote{Throughout this paper, $Dg(x)$ denotes the Jacobian matrix of $g$ at $x$.}
\begin{enumerate}[I)]
\item $u^g(x,v)$ is a well-defined continuous function on $\tilde{\Omega}^2$ and $\succsim^g$ is a complete, p-transitive, continuous, and monotone binary relation on $\tilde{\Omega}$.

\item $x\in f^{\succsim^g}(g(x),g(x)\cdot x)$ for all $x\in \tilde{\Omega}$ if and only if $w^TDg(x)w\le 0$ for all $x\in \tilde{\Omega}$ and $w\in \mathbb{R}^n$ such that $w\cdot g(x)=0$. Moreover, if $w^TDg(x)w<0$ for all $x\in \tilde{\Omega}$ and $w\in \mathbb{R}^n$ such that $w\neq 0$ and $w\cdot g(x)=0$, then $f^{\succsim^g}(p,m)$ contains at most one element, and if $f^{\succsim^g}(p,m)\neq \emptyset$, then $p$ is proportional to $g(x)$ for $x=f^{\succsim^g}(p,m)$.

\item Consider a CoD $f:P\to \tilde{\Omega}$ such that $g$ is an inverse demand function of $f$, and suppose that $f=f^{\succsim^g}$. Then, the following statements are equivalent.
\begin{enumerate}[i)]
\item $f$ is a demand function.

\item $\succsim^g$ is transitive.

\item For every $v\in \tilde{\Omega}$, $u^g_v:x\mapsto u^g(x,v)$ is $C^k$ and there exists $\lambda:\tilde{\Omega}\to \mathbb{R}$ such that $\nabla u^g_v(x)=\lambda(x)g(x)$ for all $x\in \tilde{\Omega}$.

\item For every $v\in \tilde{\Omega}$, $u^g_v:x\mapsto u^g(x,v)$ is $C^k$ and represents $\succsim^g$.

\item $g(x)$ satisfies (\ref{JACOBI}).
\end{enumerate}
\end{enumerate}

We provide the proof of this theorem in the appendix. Applying this theorem to Gale's example, we can present the following arguments. First, recall that in Gale's example, the consumption set $\Omega$ is $\mathbb{R}^3_+$ and the domain $P=\mathbb{R}^3_{++}\times\mathbb{R}_{++}$. Let $\tilde{\Omega}=\mathbb{R}^3_{++}$ and $\tilde{P}=(f^G)^{-1}(\tilde{\Omega})$. In the appendix, we check that $\tilde{P}=\{(mg(x),m)|x\in \tilde{\Omega},m>0\}$, where $g(x)$ is given by (\ref{GALE2}).\footnote{See Subsection A.6.} Define $\tilde{f}:\tilde{P}\to \tilde{\Omega}$ as the restriction of $f^G$ to $\tilde{P}$. Then, $\tilde{f}$ is also a CoD, where the consumption set is $\tilde{\Omega}$ and the domain is $\tilde{P}$.

Second, suppose that $f^G$ is a demand function. Then, $f^G=f^{\succsim}$ for some weak order $\succsim$ on $\Omega$. Let $\succsim'=\succsim\cap (\tilde{\Omega})^2$: that is, $\succsim'$ is the restriction of $\succsim$ to $\tilde{\Omega}$. Then, $\succsim'$ is a weak order on $\tilde{\Omega}$ and $\tilde{f}=f^{\succsim'}$, which implies that $\tilde{f}$ is a demand function.

Third, recall the function $g$ defined in (\ref{GALE2}). We found that
\[x=f^G(g(x),g(x)\cdot x)=\tilde{f}(g(x),g(x)\cdot x)\]
for all $x\in \tilde{\Omega}$, which implies that $g(x)$ is an inverse demand function of $\tilde{f}$. In the appendix, we show that $w^TDg(x)w<0$ for all $x\in \mathbb{R}^3_{++}$ and $w\in\mathbb{R}^3$ such that $w\neq 0$ and $w\cdot g(x)=0$.\footnote{See Subsection A.5.} By Theorem 4, $\tilde{f}=f^{\succsim^g}$ and $g(x)$ satisfies (\ref{JACOBI}) if and only if $\tilde{f}$ is a demand function.\footnote{Note that, the claim $\tilde{f}=f^{\succsim^g}$ means that $\tilde{f}(p,m)=f^{\succsim^g}(p,m)$ for all $(p,m)\in \tilde{P}$ and $f^{\succsim^g}(p,m)=\emptyset$ if $(p,m)\notin \tilde{P}$. Therefore, we needed to verify that $\tilde{P}=\{(mg(x),m)|x\in \tilde{\Omega},m>0\}$ in the above argument.} Because $g(x)$ defined in (\ref{GALE2}) violates (\ref{JACOBI}), we have that $\tilde{f}$ is not a demand function.

In conclusion, we obtain two results. 1) if $f^G$ is a demand function, then $\tilde{f}$ is also a demand function. 2) $\tilde{f}$ is not a demand function. Therefore, by the contrapositive of 1), $f^G$ is also not a demand function. This is the third proof that Gale's example is not a demand function.

\vspace{12pt}
\noindent
{\bf Notes on Theorem 4}. This result is deeply related to the ``open cycle theory'' in Pareto (1906b). In 1906, Pareto published his famous monograph called ``Manuale'' (Pareto, 1906a). It is said that Volterra wrote a review of this ``Manuale'' (Volterra, 1906), and pointed out a mathematical error in Pareto's discussion of the problem of consumer choice. In fact, there are various arguments about this story, and some say that Volterra criticized Pareto for not writing (\ref{JACOBI}), whereas others say that Volterra was not actually criticizing Pareto, but rather referring to another work of Pareto and recommending that it be incorporated into the book.\footnote{If the reader can read Japanese, we recommend Suda (2007), which includes a very detailed survey on this problem.} In any case, Pareto's response to Volterra's review was to try to construct the consumer theory for the case in which (\ref{JACOBI}) does not hold (Pareto, 1906b), and eventually, this paper was incorporated into the mathematical appendix of the French edition of his book (Pareto, 1909), known as ``Manuel''.

We explain why this theory is called ``open cycle theory'' using ``three-sided tower'' arguments due to Samuelson (1950).\footnote{We note that Samuelson consistently argued that Pareto's argument was wrong. Although Pareto connects this problem with the order of consumption, Samuelson did not think that this explanation is correct.} Samuelson explained what occurs when the integrability condition (\ref{JACOBI}) is violated using three linearly independent vectors $x,y,z$. First, consider the plane spanned by $x,y$, and draw the indifference curve passing through $x$. This curve intersects the straight line passing through $0$ and $y$ only once, and this point can be written as $ay$ for some $a>0$. Similarly, consider the plane spanned by $y,z$, and draw the indifference curve passing through $ay$. Then, this curve intersects the straight line passing through $0$ and $z$ only once, and this point can be written as $bz$ for some $b>0$. Finally, consider the plane spanned by $z,x$, and draw the indifference curve passing through $bz$. Then, this curve intersects the straight line passing through $0$ and $x$ only once, and this point can be written as $cx$ for some $c>0$. Samuelson said that $c=1$ must hold if (\ref{JACOBI}) holds, but if (\ref{JACOBI}) does not hold, then $c\neq 1$ in some cases. In other words, if (\ref{JACOBI}) is violated, then the above cycle constructed from three indifference curves is not necessarily a closed curve. Pareto's theory tried to treat the case in which the cycle constructed using indifference curves does not become a closed curve, and thus this theory is called ``open cycle theory''.

By the way, Samuelson called the consumer who violates (\ref{JACOBI}) ``a man that can be easily cheated.'' To illustrate why, consider the above example with $c<1$. In this case, when the consumer is offered to exchange $x$ for $ay$, then he/she accepts this exchange because these are indifferent. Next, if he/she is offered to exchange $ay$ for $bz$, then he/she also accepts this because these are indifferent. Finally, when someone offers him/her to exchange $bz$ for $cx$, he/she accepts this because these are indifferent. As a result of these exchanges, his/her initial consumption vector $x$ is reduced to $cx$. Samuelson described such an individual as ``easily cheated.'' Of course, a similar exchange can be done in the case $c>1$.

This discussion can be further understood using Theorem 4. First, the ODE (\ref{IND}) is intentionally designed so that the right-hand side is orthogonal to $g(y(t))$. In microeconomics, the budget hyperplane is orthogonal to the price vector, and at the optimal consumption plan, the indifference hypersurface is tangent to the budget hyperplane. Therefore, the supporting hyperplane of the indifference hypersurface is orthogonal to the price vector at this point. Because $g(y(t))$ denotes the price vector under which $y(t)$ is optimal, the trajectory of the curve $y(t;x,v)$ can be seen as the indifference curve passing through $x$ in the plane spanned by $x,v$. Hence, translating Samuelson's arguments into our symbols,
\[ay=y(t(x,y);x,y),\ bz=y(t(ay,z);y,z),\ cx=y(t(bz,x);z,x).\]
By an easy calculation, we obtain that
\[a=u^g(x,y),\ b=u^g(ay,z),\ c=u^g(bz,x).\]
It is easy to verify that
\[x\sim^gay,\ ay\sim^gbz,\ bz\sim^gcx.\]
Therefore, if $\succsim^g$ is transitive, then $c=1$. However, if $\succsim^g$ is not transitive, then there exist $x,y,z$ such that $c\neq 1$. Hence, Theorem 4 can be viewed as the result of a mathematical expression of Samuelson's explanation.

Probably, Gale had this Samuelson's argument in mind and thus focused on the fact that $g(x)$ in (\ref{GALE2}) does not satisfy (\ref{JACOBI}). In this sense, the proof in this subsection can be considered to be the closest to the argument that Gale originally had in mind.\footnote{However, it cannot be assumed that Gale himself knew or substantially proved Theorem 4. This can be seen in the proof of Theorem 4. As we see in the appendix, the proof of Theorem 4 is extremely long and difficult. If Gale had proved such a result, we cannot believe that he would not have left a record of it anywhere. In this regard, although it is possible to prove that $f^G$ is not a demand function using Gale's own idea, it is inconceivable that Gale would have done so.}

Samuelson discussed using three linearly independent vectors, and he did not have a problem with the fact that an indifference curve can be drawn. In other words, he thought that such a problem would not occur for the case in which $n=2$. In this connection, we mention two facts. First, (\ref{JACOBI}) automatically holds if any two of $i,j,k$ coincide, and therefore it holds unconditionally when $n=2$. Second, Rose (1958) showed that the weak axiom implies the strong axiom when $n=2$. Therefore, in light of Theorem 1, there is no CoD that satisfies the weak axiom but is not a demand function. Indeed, Gale's example treated the case in which $n=3$, which is an essential assumption.

We must mention Hurwicz and Richter (1979a, b). They considered an axiom called {\bf Ville's axiom of revealed preference} (or simply, {\bf Ville's axiom}) for a function $g:\mathbb{R}^n_{++}\to\mathbb{R}^n_{++}$. A function $g:\mathbb{R}^n_{++}\to \mathbb{R}^n_{++}$ is said to satisfy Ville's axiom if there is no piecewise $C^1$ closed curve $x:[0,T]\to \mathbb{R}^n_{++}$ such that
\begin{equation}\label{VILLE}
g(x(t))\cdot \dot{x}(t)>0
\end{equation}
for almost all $t\in [0,T]$. They showed that when $g$ is $C^1$, $g$ satisfies this axiom if and only if $g$ satisfies (\ref{JACOBI}). For a modern proof of this result, see Hosoya (2019). Because $g(x)$ defined by (\ref{GALE2}) violates (\ref{JACOBI}), this $g(x)$ also violates Ville's axiom. Therefore, there exists a closed curve $x(t)$ that satisfies (\ref{VILLE}) for almost all $t$.

If $f^G=f^u$ for some $C^1$ utility function $u$, then by Lagrange's multiplier rule, $\nabla u(x)=\lambda(x)g(x)$ for all $x$. Therefore,
\[\frac{d}{dt}u(x(t))>0\]
for almost all $t\in [0,T]$, which contradicts $x(0)=x(T)$. Ville's axiom is built on this idea, and it is worth noting that this idea and the ``open cycle'' argument are very similar. In fact, we can easily show the following result.

\vspace{12pt}
\noindent
{\bf Theorem 5}. Suppose that $g:\tilde{\Omega}\to \mathbb{R}^n_{++}$ is a $C^1$ function. Then, $g$ satisfies Ville's axiom if and only if $\succsim^g$ is transitive.

\vspace{12pt}
We present a proof in the appendix.

Finally, we mention a fact. We have already shown that the restriction $\tilde{f}$ of Gale's example is a semi-demand function. Actually, Hosoya (2021b) showed that any CoD that satisfies Walras' law and the weak axiom is a semi-demand function corresponds to some complete binary relation. Hence, Gale's example itself is also a semi-demand function. We think that Gale's example is actually a semi-demand function corresponds to a complete, p-transitive, and continuous binary relation, and such a binary relation can be constructed in the same manner as Theorem 4. However, there are several technical difficulties, and thus it is an open problem.

\section{Conclusion}
We have scrutinized Gale's paper written in 1960 to see how far Gale had shown. As a result, we found that Gale showed that his constructed CoD satisfies the weak axiom of revealed preference, and that the corresponding inverse demand function does not satisfy Jacobi's integrability condition, but he did not show that this CoD is not a demand function. Next, we proved that this CoD is not a demand function in three ways. The first is the proof that Gale was able to construct in 1960, the second is the simplest modern proof, and the third is the proof constructed in the direction Gale originally intended.

We paid particular attention to the first proof, and found that, although what Gale showed in his paper was insufficient, we could easily confirm Gale's claim from what he showed. Thus, it is fair to say that the credit for discovering a CoD that is not a demand function while satisfying the weak axiom of revealed preference belongs to Gale.

\appendix
\section{Proofs of Results}
\subsection{Proof of the Well-Definedness of $\bar{p}$ in Gale's Example}
In this subsection, we prove the well-definedness of $\bar{p}$. 

Recall the definition of $\bar{p}$. For $p\in \mathbb{R}^3_{++}$, the definition of $\bar{p}$ is as follows.
\begin{enumerate}[I)]
\item If $p\in C$, define $\bar{p}=p$.

\item Suppose that for some $(i,j,k)\in \{(1,2,3),(2,3,1),(3,1,2)\}$, $-3p_i+4p_j\le 0$, and $-3p_j+4p_k\le 0$. By our previous argument, we have that $-3p_k+4p_i>0$. In this case, define $\bar{p}_i=\frac{16}{9}p_k$, $\bar{p}_j=\frac{4}{3}p_k$, and $\bar{p}_k=p_k$.

\item Suppose that for some $(i,j,k)\in \{(1,2,3),(2,3,1),(3,1,2)\}$, $-3p_i+4p_j\le 0$, $-3p_j+4p_k\ge 0$, and $-3p_k+4p_i\ge 0$. We separate this case into two subcases.
\begin{enumerate}[i)]
\item If $16p_j-9p_k\ge 0$, then define $\bar{p}_i=\frac{4}{3}p_j$, $\bar{p}_j=p_j$, and $\bar{p}_k=p_k$.

\item If $16p_j-9p_k\le 0$, then define $\bar{p}_i=\frac{4}{3}p_j$, $\bar{p}_j=p_j$, and $\bar{p}_k=\frac{16}{9}p_j$.
\end{enumerate}
\end{enumerate}

Our problem is the existence of $p$ for which multiple definitions are applicable. For example, if $p=(4,3,4)$, then $-3p_1+4p_2=0$, $-3p_2+4p_3>0$ and $-3p_3+4p_1>0$, which implies that I) and III)-i) are applicable for this $p$. We need to show that, in such a case, both definition in I) and III)-i) lead to the same $\bar{p}$.

We separate the proof into seven cases. Let $I=\{(1,2,3),(2,3,1),(3,1,2)\}$. Recall that if $-3p_i+4p_j\le 0$ and $-3p_j+4p_k\le 0$, then $-3p_k+4p_i>0$.

Define $C_1=C$ and
\begin{align*}
C_2=\{p\in \mathbb{R}^3_{++}|&-3p_i+4p_j\le 0,\ -3p_j+4p_k\le 0\mbox{ for some }(i,j,k)\in I\},\\
C_3=\{p\in \mathbb{R}^3_{++}|&-3p_i+4p_j\le 0,\ -3p_j+4p_k\ge 0,\\
&~-3p_k+4p_i\ge 0,\ 16p_j-9p_k\ge 0\mbox{ for some }(i,j,k)\in I\},\\
C_4=\{p\in \mathbb{R}^3_{++}|&-3p_i+4p_j\le 0,\ -3p_j+4p_k\ge 0,\\
&~-3p_k+4p_i\ge 0,\ 16p_j-9p_k\le 0\mbox{ for some }(i,j,k)\in I\}.
\end{align*}
Then, $p\in C_1$ if and only if I) is applicable, $p\in C_2$ if and only if II) is applicable, $p\in C_3$ if and only if III)-i) is applicable, and $p\in C_4$ if and only if III)-ii) is applicable.

\vspace{12pt}
\noindent
{\bf Case 1}. $-3p_i+4p_j>0,\ -3p_j+4p_k>0,\ -3p_k+4p_i>0$.

In this case, $p\in C_1$ but $p\notin C_2\cup C_3\cup C_4$. Thus, only definition I) is applicable, and thus $\bar{p}=p$.

\vspace{12pt}
\noindent
{\bf Case 2}. $-3p_i+4p_j=0,\ -3p_j+4p_k>0,\ -3p_k+4p_i>0$.

In this case, $-9p_k+16p_j>0$, and thus $p\in C_1\cap C_3$ but $p\notin C_2\cup C_4$. Hence, definitions I) and III)-i) are applicable, and in both definitions, $\bar{p}=p$.

\vspace{12pt}
\noindent
{\bf Case 3}. $-3p_i+4p_j<0,\ -3p_j+4p_k>0,\ -3p_k+4p_i>0$.

In this case, if $-9p_k+16p_j>0$, then $p\in C_3$ but $p\notin C_1\cup C_2\cup C_4$. If $-9p_k+16p_j<0$, then $p\in C_4$ but $p\notin C_1\cup C_2\cup C_3$. In both cases, $\bar{p}$ is well-defined. If $-9p_k+16p_j=0$, then $p\in C_3\cap C_4$ and thus definitions III)-i) and III)-ii) are applicable, and in both definitions, $\bar{p}$ are the same.

\vspace{12pt}
\noindent
{\bf Case 4}. $-3p_i+4p_j=0,\ -3p_j+4p_k=0,\ -3p_k+4p_i>0$.

Define $i^*=j, j^*=k, k^*=i$. In this case, $-9p_k+16p_j>0$ and $-9p_{k^*}+16p_{j^*}=0$. This implies that $p\in C_1\cap C_2\cap C_3\cap C_4$, and thus definitions I), II), III)-i), and III)-ii) are applicable. And in all cases, $\bar{p}=p$.

\vspace{12pt}
\noindent
{\bf Case 5}. $-3p_i+4p_j=0,\ -3p_j+4p_k<0,\ -3p_k+4p_i>0$.

Define $i^*=j, j^*=k, k^*=i$. Then, $-9p_{k^*}+16p_{j^*}<0$, and thus $p\in C_2\cap C_4$ but $p\notin C_1\cup C_3$. Hence, definitions II) and III)-ii) are applicable, and in both cases, $\bar{p}$ are the same.

\vspace{12pt}
\noindent
{\bf Case 6}. $-3p_i+4p_j<0,\ -3p_j+4p_k=0,\ -3p_k+4p_i>0$.

In this case, $-9p_k+16p_j>0$, and thus $p\in C_2\cap C_3$ but $p\notin C_1\cup C_4$. Hence, definitions II) and III)-i) are applicable, and in both cases, $\bar{p}$ are the same.

\vspace{12pt}
\noindent
{\bf Case 7}. $-3p_i+4p_j<0,\ -3p_j+4p_k<0,\ -3p_k+4p_i>0$.

In this case, $p\in C_2$ but $p\notin C_1\cup C_3\cup C_4$, and thus only definition II) is applicable, and thus $\bar{p}$ is defined.

\vspace{12pt}
Hence, in all cases $\bar{p}$ is well-defined.

Using this argument, we can easily show that the mapping $p\mapsto \bar{p}$ is continuous. Actually, the restriction of this mapping to $C_i$ is trivially continuous, and if $p\in C_i\cap C_j$, then the value of these functions coincide. The continuity of this mapping immediately follows from this fact.

\subsection{Proof of the Basic Property on the Transitive Closure}
In this subsection, we prove the following.

\vspace{12pt}
\noindent
{\bf Lemma 1}. Let $X\neq \emptyset$ and suppose that $R$ is a binary relation on $X$. Let $R^*$ be the transitive closure of $R$. Then, $xR^*y$ if and only if there exists a finite sequence $x_1,...,x_k$ such that $x_1=x,\ x_k=y$ and $x_iRx_{i+1}$ for all $i\in \{1,...,k-1\}$.

\vspace{12pt}
\noindent
{\bf Proof}. Define $R^+$ as the set of all $(x,y)\in X^2$ such that there exists a finite sequence $x_1,...,x_k$ such that $x_1=x,\ x_k=y$ and $x_iRx_{i+1}$ for all $i\in \{1,...,k-1\}$. It suffices to show that $R^+=R^*$.

It is obvious that $R^+$ is transitive, and thus $R^*\subset R^+$.

Conversely, suppose that $xR^+y$. Then, there exists a finite sequence $x_1,...,x_k$ such that $x_1=x,\ x_k=y$ and $x_iRx_{i+1}$ for all $i\in \{1,...,k-1\}$. Because $R\subset R^*$, we have that $x_iR^*x_{i+1}$ for all $i\in \{1,...,k-1\}$. Since $R^*$ is transitive, we have that $xR^*y$, which implies that $R^+\subset R^*$. This completes the proof. $\blacksquare$

\subsection{Proof of Theorem 1}
Suppose that $f=f^{\succsim}$ for some weak order $\succsim$. We have already shown that $x\succ_ry$ implies that $x\succ y$. Hence, Lemma 1 and the transitivity of $\succ$ mean that $x\succ_{ir}y$ implies that $x\succ y$. Hence,
\[x\succ_{ir}y\Rightarrow x\succ y\Rightarrow y\not\succ x\Rightarrow y\not\succ_{ir}x,\]
which implies that $\succ_{ir}$ is asymmetric, and thus $f$ satisfies the strong axiom.

To prove the converse relationship, we need the following lemma.

\vspace{12pt}
\noindent
{\bf Lemma 2} (Szpilrajn's extension theorem). Suppose that $X$ is a nonempty set and $\succ$ is a strong order on $X$. Then, there exists an antisymmetric weak order $\succsim^*$ that includes $\succ$.\footnote{This theorem was shown by Szpilrajn (1930). For another modern proof, see Chambers and Echenique (2016).}

\vspace{12pt}
\noindent
{\bf Proof}. Let $L$ be the set of all strong orders on $X$ that includes $\succ$. Define
\[\succeq=\{(\succ_1,\succ_2)\in L^2|\succ_2\subset \succ_1\}.\]
Then, $\succeq$ is a reflexive and transitive binary relation on $L$, and clearly every chain $C$ of $\succeq$ has an upper bound. By Zorn's lemma, there exists a maximal element $\succ^*\subset L$ with respect to $\succeq$. Define
\[\succsim^*=\succ^*\cup \{(x,x)|x\in X\}.\]
Clearly $\succsim^*$ is antisymmetric, and it is easy to show that $\succsim^*$ is transitive. It suffices to show that $\succsim^*$ is complete.

Suppose not. Then, there exists $x,y\in X$ such that $x\not\succsim^*y$ and $y\not\succsim^*x$. By the definition of $\succsim^*$, we have that $x\neq y$. Define
\[\succ^+=\succ^*\cup\{(z,w)\in X^2|z\succsim^*x,\ y\succsim^*w\}.\]
We first show that $\succ^+$ is asymmetric. Suppose not. Then, there exist $z,w\in X$ such that $z\succ^+w$ and $w\succ^+z$. If $z\not\succ^*w$ and $w\not\succ^*z$, then $z\succsim^*x$ and $y\succsim^*z$. This implies that $y\succsim^*x$, which is a contradiction. Therefore, either $z\succ^*w$ or $w\succ^*z$. Without loss of generality, we assume that $w\succ^*z$. Then, $z\not\succ^*w$, and thus $z\succsim^*x$ and $y\succsim^*w$. Because $w\succ^*z$, we have that $w\succsim^*z$, and by the transitivity of $\succsim^*$, $y\succsim^*x$, which is a contradiction.

Second, we show that $\succ^+$ is transitive. Suppose that $z\succ^+w$ and $w\succ^+v$. If $z\succ^*w$ and $w\succ^*v$, then $z\succ^*v$, and thus $z\succ^+v$. Suppose that $z\not\succ^*w$. Then, $z\succsim^*x$ and $y\succsim^*w$. If $w\not\succ^*v$, then $w\succsim^*x$. This implies that $y\succsim^*x$, which is a contradiction. Hence, we have that $w\succ^*v$. This implies that $y\succsim^*v$, and thus $z\succ^+w$. Next, suppose that $z\succ^*w$ and $w\not\succ^*v$. Then, $w\succsim^*x$ and $y\succsim^*v$. Because $z\succ^*w$, we have that $z\succsim^*x$, which implies that $z\succ^+v$. Thus, in any case, we have that $z\succ^+v$, which implies that $\succ^+$ is transitive.

Therefore, $\succ^+$ is a strong order that includes $\succ^*$. Because $x\succ^+y$ and $x\not\succ^*y$, we have that $\succ^*\subsetneq \succ^+$. This implies that $\succ^*$ is not maximal with respect to $\succeq$, which is a contradiction. This completes the proof. $\blacksquare$

\vspace{12pt}
Now, suppose that $f$ satisfies the strong axiom. Then, $\succ_{ir}$ is a strong order on $\Omega$, and thus there exists an antisymmetric weak order $\succsim^*$ on $\Omega$ that includes $\succ_{ir}$. Suppose that $x=f(p,m)$ and $y\in \Delta(p,m)$. If $x=y$, then $x\succsim^*y$. If $x\neq y$, then $x\succ_{ir}y$, and thus $x\succsim^*y$. Moreover, because $\succsim^*$ is antisymmetric, $y\not\succsim^*x$. This implies that $x=f^{\succsim^*}(p,m)$, and thus $f=f^{\succsim^*}$. Hence, $f$ is a demand function. This completes the proof of Theorem 1. $\blacksquare$

\subsection{Proof of Theorem 3}
Choose any $p,q$ and $t\in [0,1]$, and define $r=(1-t)p+tq$. Choose any $\varepsilon>0$. Then, there exists $y\in \Omega$ such that $y\succsim x$ and $r\cdot y\le E^x(r)+\varepsilon$. Thus,
\[E^x(r)+\varepsilon\ge r\cdot y=(1-t)p\cdot y+tq\cdot y\ge (1-t)E^x(p)+tE^x(q),\]
which implies that $E^x$ is concave. Because any concave function defined on an open set is continuous, $E^x$ is continuous.

In the reminder of the proof, we assume that $f$ satisfies Walras' law and $x\in R(f)$. Suppose that $x=f(p^*,m^*)$. Choose any $p\in \mathbb{R}^n_{++}$. Then, there exists $\varepsilon>0$ such that $p^*\cdot y<m^*$ for all $y\in \Delta(p,\varepsilon)$. In this case, $x\succ y$ for all $y\in \Delta(p,\varepsilon)$, and thus $E^x(p)>\varepsilon$. Hence, we have that $E^x(p)>0$, and thus 1) holds.

Next, suppose that $y\in \Delta(p^*,m^*)$ and $x\neq y$. Then, $x\succ y$. By the contrapositive of this, we have that if $y\succsim x$, then either $y=x$ or $p^*\cdot y>m^*$, which implies that $E^x(p^*)=m^*$. Hence, 2) holds.

Third, choose any $p\in\mathbb{R}^n_{++}$. Define
\[D_{i,+}E^x(p)=\lim_{h\downarrow 0}\frac{E^x(p+he_i)-E^x(p)}{h},\ D_{i,-}E^x(p)=\lim_{h\uparrow 0}\frac{E^x(p+he_i)-E^x(p)}{h},\]
where $e_i$ denotes the $i$-th unit vector. Because $E^x$ is concave, both limits are defined, and
\[D_{i,+}E^x(p)\le D_{i,-}E^x(p).\]
For any $q\in \mathbb{R}^n_{++}$ and $\varepsilon>0$, define $X(q)=f(q,E^x(q))$ and $X^{\varepsilon}(q)=f(q,E^x(q)+\varepsilon)$. By definition of $E^x(q)$, there exists $y\in \Omega$ such that $y\succsim x$ and $q\cdot y\le E^x(q)+\varepsilon$. Because $X^{\varepsilon}(q)\succsim y$, we have that $X^{\varepsilon}(q)\succsim x$, and thus $p\cdot X^{\varepsilon}(q)\ge E^x(p)$. Because $f$ is continuous, letting $\varepsilon\to 0$, we obtain
\[p\cdot X(q)\ge E^x(p)=p\cdot X(p).\]
Therefore, if we set $q=p+he_i$, then
\begin{align*}
E^x(q)-E^x(p)=&~q\cdot X(q)-p\cdot X(p)\\
=&~p\cdot (X(q)-X(p))+hX_i(q)\\
\ge&~hX_i(q).
\end{align*}
By the above inequality,
\[D_{i,+}E^x(p)\ge f_i(p,E^x(p))\ge D_{i,-}E^x(p),\]
and thus,
\[\frac{\partial E^x}{\partial p_i}(p)=f_i(p,E^x(p)),\]
which implies that (\ref{Shephard}) holds. This completes the proof. $\blacksquare$

\subsection{Property of the Inverse Demand Function}
In the main text, we claimed several property of the inverse demand function $g(x)$ defined in (\ref{GALE2}). In this subsection, we present proofs of these facts.

First, we present a lemma.

\vspace{12pt}
\noindent
{\bf Lemma 3}. Suppose that $U\subset \mathbb{R}^n$ is open, $g:U\to \mathbb{R}^n\setminus\{0\}$ is $C^1$, and $\lambda:U\to \mathbb{R}_{++}$ is also $C^1$. Define $h(x)=\lambda(x)g(x)$. Then, the following holds.
\begin{enumerate}[1)]
\item $g(x)$ satisfies (\ref{JACOBI}) if and only if $h(x)$ satisfies (\ref{JACOBI}).

\item $w^TDg(x)w\le 0$ for all $x\in U$ and $w\in \mathbb{R}^n$ such that $w\cdot g(x)=0$ if and only if $w^TDh(x)w\le 0$ for all $x\in U$ and $w\in \mathbb{R}^n$ such that $w\cdot h(x)=0$.

\item $w^TDg(x)w<0$ for all $x\in U$ and $w\in\mathbb{R}^n$ such that $w\neq 0$ and $w\cdot g(x)=0$ if and only if $w^TDh(x)w<0$ for all $x\in U$ and $w\in \mathbb{R}^n$ such that $w\neq 0$ and $w\cdot h(x)=0$.
\end{enumerate}

\vspace{12pt}
\noindent
{\bf Proof}. By Leibniz's rule,
\begin{align*}
&~h_i\left(\frac{\partial h_j}{\partial x_k}-\frac{\partial h_k}{\partial x_j}\right)+h_j\left(\frac{\partial h_k}{\partial x_i}-\frac{\partial h_i}{\partial x_k}\right)+h_k\left(\frac{\partial h_i}{\partial x_j}-\frac{\partial h_j}{\partial x_i}\right)\\
=&~\lambda^2\left[g_i\left(\frac{\partial g_j}{\partial x_k}-\frac{\partial g_k}{\partial x_j}\right)+g_j\left(\frac{\partial g_k}{\partial x_i}-\frac{\partial g_i}{\partial x_k}\right)+g_k\left(\frac{\partial g_i}{\partial x_j}-\frac{\partial g_j}{\partial x_i}\right)\right]\\
&~+\lambda\left[g_i\left(g_j\frac{\partial \lambda}{\partial x_k}-g_k\frac{\partial \lambda}{\partial x_j}\right)+g_j\left(g_k\frac{\partial \lambda}{\partial x_i}-g_i\frac{\partial \lambda}{\partial x_k}\right)+g_k\left(g_i\frac{\partial \lambda}{\partial x_j}-g_j\frac{\partial \lambda}{\partial x_i}\right)\right]\\
=&~\lambda^2\left[g_i\left(\frac{\partial g_j}{\partial x_k}-\frac{\partial g_k}{\partial x_j}\right)+g_j\left(\frac{\partial g_k}{\partial x_i}-\frac{\partial g_i}{\partial x_k}\right)+g_k\left(\frac{\partial g_i}{\partial x_j}-\frac{\partial g_j}{\partial x_i}\right)\right],
\end{align*}
and thus, $h(x)$ satisfies (\ref{JACOBI}) if and only if $g(x)$ satisfies (\ref{JACOBI}). Hence, 1) holds.

Next, we note that $w\cdot g(x)=0$ if and only if $w\cdot h(x)=0$. If $w\cdot g(x)=0$, then
\[w^TDh(x)w=w^Tg(x)D\lambda(x)w+\lambda(x)w^TDg(x)w=\lambda(x)w^TDg(x)w,\]
and thus, 2) and 3) hold. This completes the proof. $\blacksquare$

\vspace{12pt}
We show that $g(x)$ defined in (\ref{GALE2}) violates (\ref{JACOBI}). Choosing $\lambda(x)=37(x^TBx)$ and applying Lemma 3, we have that $g(x)$ satisfies (\ref{JACOBI}) if and only if
\begin{equation}\label{GALE3}
h(x)=\begin{pmatrix}
9 & 12 & 16\\
16 & 9 & 12\\
12 & 16 & 9
\end{pmatrix}\begin{pmatrix}
x_1\\
x_2\\
x_3
\end{pmatrix}
\end{equation}
satisfies (\ref{JACOBI}). However, if $x_1=x_2=x_3=1$, then
\begin{align*}
&~h_1\left(\frac{\partial h_2}{\partial x_3}-\frac{\partial h_3}{\partial x_2}\right)+h_2\left(\frac{\partial h_3}{\partial x_1}-\frac{\partial h_1}{\partial x_3}\right)+h_3\left(\frac{\partial h_1}{\partial x_2}-\frac{\partial h_2}{\partial x_1}\right)\\
=&~37(12-16)+37(12-16)+37(12-16)=-444\neq 0,
\end{align*}
and thus $h(x)$ violates (\ref{JACOBI}).

Define
\[C=\begin{pmatrix}
9 & 14 & 14\\
14 & 9 & 14\\
14 & 14 & 9
\end{pmatrix}.\]
Then, $w^TDh(x)w=w^TCw$, where $h(x)$ is defined in (\ref{GALE3}). Let $t=\frac{h_1(x)}{h_2(x)}$. Then, $\frac{9}{16}\le t\le \frac{4}{3}$, and thus
\[\begin{vmatrix}
9 & 14 & h_1(x)\\
14 & 9 & h_2(x)\\
h_1(x) & h_2(x) & 0
\end{vmatrix}=(h_2(x))^2(-9t^2+28t-9)>0.
\]
Next, let $s_1=\frac{h_1(x)}{h_3(x)}$ and $s_2=\frac{h_2(x)}{h_3(x)}$. Then, $\frac{3}{4}\le s_1\le \frac{16}{9}$, $\frac{9}{16}\le s_2\le \frac{4}{3}$, and $\frac{9}{16}\le \frac{s_1}{s_2}\le \frac{4}{3}$, and thus
\begin{align*}
&~\begin{vmatrix}
9 & 14 & 14 & h_1(x)\\
14 & 9 & 14 & h_2(x)\\
14 & 14 & 9 & h_3(x)\\
h_1(x) & h_2(x) & h_3(x) & 0
\end{vmatrix}\\
=&~(h_3(x))^2[115s_1^2-140s_1s_2+115s_2^2-140s_1-140s_2+115]<0.
\end{align*}
By Theorem 5 of Debreu (1952), we have that $w^TCw<0$ for all $w\in \mathbb{R}$ such that $w\neq 0$ and $w\cdot h(x)=0$. Therefore, by Lemma 3, $w^TDg(x)w<0$ for all $x\in \mathbb{R}^3_{++}$ and $w\in \mathbb{R}^3$ such that $w\neq 0$ and $w\cdot g(x)=0$.

\subsection{Proof of a Fact on the Domain of Gale's Example}
In Subsection 4.3, we first introduce Theorem 4 and then provide a proof that $f^G$ is not a demand function. In this proof, we defined $\tilde{\Omega}=\mathbb{R}^3_{++}$ and $\tilde{P}=(f^G)^{-1}(\tilde{\Omega})$, and claimed that $\tilde{P}=\{(mg(x),m)|x\in \tilde{\Omega},\ m>0\}$, where $g(x)$ is defined by (\ref{GALE2}). In this subsection, we prove this claim rigorously.

First, we present three lemmas.

\vspace{12pt}
\noindent
{\bf Lemma 4}. Suppose that $U\subset \mathbb{R}^n$ is open, and $g:U\to \mathbb{R}^n\setminus \{0\}$ is $C^1$. Suppose also that $g_n(x)\neq 0$. Then,
\begin{equation}\label{JACOBI2}
g_i\left(\frac{\partial g_j}{\partial x_n}-\frac{\partial g_n}{\partial x_j}\right)+g_j\left(\frac{\partial g_n}{\partial x_i}-\frac{\partial g_i}{\partial x_n}\right)+g_n\left(\frac{\partial g_i}{\partial x_j}-\frac{\partial g_j}{\partial x_i}\right)=0
\end{equation}
for all $i,j\in \{1,...,n-1\}$ if and only if $g$ satisfies (\ref{JACOBI}) for all $i,j,k\in \{1,...,n\}$.

\vspace{12pt}
\noindent
{\bf Proof}. Clearly, (\ref{JACOBI}) implies (\ref{JACOBI2}). Conversely, suppose that (\ref{JACOBI2}) holds for all $i,j\in \{1,...,n-1\}$. Choose any $i,j,k\in \{1,...,n\}$. If two or three of them are $n$, then trivially (\ref{JACOBI}) holds. If one of them is $n$, then by (\ref{JACOBI2}), (\ref{JACOBI}) holds. Therefore, without loss of generality, we can assume that $i,j,k\in \{1,...,n-1\}$. Then,
\[g_i\left(\frac{\partial g_j}{\partial x_n}-\frac{\partial g_n}{\partial x_j}\right)+g_j\left(\frac{\partial g_n}{\partial x_i}-\frac{\partial g_i}{\partial x_n}\right)+g_n\left(\frac{\partial g_i}{\partial x_j}-\frac{\partial g_j}{\partial x_i}\right)=0,\]
\[g_j\left(\frac{\partial g_k}{\partial x_n}-\frac{\partial g_n}{\partial x_k}\right)+g_k\left(\frac{\partial g_n}{\partial x_j}-\frac{\partial g_j}{\partial x_n}\right)+g_n\left(\frac{\partial g_j}{\partial x_k}-\frac{\partial g_k}{\partial x_j}\right)=0,\]
\[g_k\left(\frac{\partial g_i}{\partial x_n}-\frac{\partial g_n}{\partial x_i}\right)+g_i\left(\frac{\partial g_n}{\partial x_k}-\frac{\partial g_k}{\partial x_n}\right)+g_n\left(\frac{\partial g_k}{\partial x_i}-\frac{\partial g_i}{\partial x_k}\right)=0.\]
Multiplying $g_k$ by the first line, $g_i$ by the second line, and $g_j$ by the third line and summing them, we obtain 
\[g_n\left[g_i\left(\frac{\partial g_j}{\partial x_k}-\frac{\partial g_k}{\partial x_j}\right)+g_j\left(\frac{\partial g_k}{\partial x_i}-\frac{\partial g_i}{\partial x_k}\right)+g_k\left(\frac{\partial g_i}{\partial x_j}-\frac{\partial g_j}{\partial x_i}\right)\right]=0,\]
and because $g_n(x)\neq 0$, (\ref{JACOBI}) holds. This completes the proof. $\blacksquare$

\vspace{12pt}
\noindent
{\bf Lemma 5}. Suppose that $U\subset \mathbb{R}^n$ is open, and $g:U\to \mathbb{R}^n\setminus \{0\}$ is $C^1$. Moreover, suppose that $g_n\equiv 1$. Define
\[a_{ij}(x)=\frac{\partial g_i}{\partial x_j}(x)-\frac{\partial g_i}{\partial x_n}(x)g_j(x),\]
and let $A_g(x)$ be the $(n-1)\times (n-1)$ matrix whose $(i,j)$-th component is $a_{ij}(x)$. Then, the following holds.\footnote{The matrix-valued function $A_g$ is sometimes called the {\bf Antonelli matrix} of $g$.}
\begin{enumerate}[1)]
\item $g(x)$ satisfies (\ref{JACOBI}) if and only if $A_g(x)$ is symmetric for all $x\in U$.

\item $w^TDg(x)w\le 0$ for all $x\in U$ and $w\in \mathbb{R}^n$ such that $w\cdot g(x)=0$ if and only if $A_g(x)$ is negative semi-definite for all $x\in U$.

\item $w^TDg(x)w<0$ for all $x\in U$ and $w\in \mathbb{R}^n$ such that $w\neq 0$ and $w\cdot g(x)=0$ if and only if $A_g(x)$ is negative definite for all $x\in U$.
\end{enumerate}

\vspace{12pt}
\noindent
{\bf Proof}. By Lemma 4, $g(x)$ satisfies (\ref{JACOBI}) if and only if $g(x)$ satisfies (\ref{JACOBI2}). Therefore, $g(x)$ satisfies (\ref{JACOBI}) if and only if for every $x\in U$ and $i,j\in \{1,...,n-1\}$,
\[g_i\left(\frac{\partial g_j}{\partial x_n}-\frac{\partial g_n}{\partial x_j}\right)+g_j\left(\frac{\partial g_n}{\partial x_i}-\frac{\partial g_i}{\partial x_n}\right)+g_n\left(\frac{\partial g_i}{\partial x_j}-\frac{\partial g_j}{\partial x_i}\right)=0.\]
Because $g_n\equiv 1$, we have that $\frac{\partial g_n}{\partial x_i}=\frac{\partial g_n}{\partial x_j}=0$, and thus the above formula can be transformed as follows:
\[g_i\frac{\partial g_j}{\partial x_n}-g_j\frac{\partial g_i}{\partial x_n}+\frac{\partial g_i}{\partial x_j}-\frac{\partial g_j}{\partial x_i}=0,\]
where the left-hand side is $a_{ij}-a_{ji}$. Therefore, (\ref{JACOBI}) is equivalent to the symmetry of $A_g$, and 1) holds.

Next, define $\hat{g}=(g_1,...,g_{n-1})$. By an easy calculation,
\[Dg(x)=\left(
\begin{array}{c|c}
A_g(x) & \frac{\partial \hat{g}}{\partial x_n}(x)\\ \hline
0^T & 0
\end{array}
\right)+\left(
\begin{array}{c|c}
\frac{\partial \hat{g}}{\partial x_n}(x)\hat{g}^T(x) & 0\\ \hline
0^T & 0
\end{array}
\right).\]
Choose any $w\in \mathbb{R}^n$ such that $w\cdot g(x)=0$, and define $\hat{w}=(w_1,...,w_{n-1})$. Then, $\hat{w}\cdot \hat{g}(x)=-w_n$, and thus
\begin{align*}
w^TDg(x)w=&~\hat{w}^TA_g(x)\hat{w}+\hat{w}^T\frac{\partial \hat{g}}{\partial x_n}(x)w_n+\hat{w}^T\frac{\partial \hat{g}}{\partial x_n}(x)\hat{g}^T(x)\hat{w}\\
=&~\hat{w}^TA_g(x)\hat{w},
\end{align*}
which implies that 2) holds. Moreover, if $w\neq 0$ and $w\cdot g(x)=0$, then $\hat{w}\neq 0$, which implies that 3) holds. This completes the proof. $\blacksquare$

\vspace{12pt}
\noindent
{\bf Lemma 6}. Suppose that $f:P\to \Omega$ is a CoD that satisfies Walras' law, and $g:\mathbb{R}^n_{++}\to \mathbb{R}^n_{++}$ is an inverse demand function of $f$ such that $g_n\equiv 1$. Let $x=f(p,m)$, and suppose that $f$ is differentiable at $(p,m)$ and $g$ is differentiable at $x$. Define $\tilde{S}_f(p,m)$ as the $(n-1)\times (n-1)$ matrix whose $(i,j)$-th component is $s_{ij}(p,m)$. Then, the matrix $A_g(x)$ is regular and $(A_g(x))^{-1}=\tilde{S}_f(p,m)$.

\vspace{12pt}
\noindent
{\bf Proof}. Define\footnote{In this proof, we frequently abbreviate variables. Note that we use the symbol $s_{i,j}$ instead of $s_{ij}$ because if $j=n-1$, then the latter expression is confusing.}
\[\hat{S}=\begin{pmatrix}
s_{1,1} & ... & s_{1,n-1} & \frac{\partial f_1}{\partial m}\\
\vdots & \ddots & \vdots & \vdots\\
s_{n,1} & ... & s_{n,n-1} & \frac{\partial f_n}{\partial m}
\end{pmatrix},\ F=\begin{pmatrix}
\frac{\partial f_1}{\partial p_1} & ... & \frac{\partial f_1}{\partial p_{n-1}} & \frac{\partial f_1}{\partial m}\\
\vdots & \ddots & \vdots & \vdots\\
\frac{\partial f_n}{\partial p_1} & ... & \frac{\partial f_n}{\partial p_{n-1}} & \frac{\partial f_n}{\partial m}
\end{pmatrix}.\]
In other words, $\hat{S}$ is the matrix whose $n$-th column vector is $D_mf(p,m)$ and $j$-th column vector is the same as that of $S_f(p,m)$ for $j\neq n$, and $F$ is the matrix whose $n$-th column vector is $D_mf(p,m)$ and $j$-th column vector is the same as that of $D_pf(p,m)$ for $j\neq n$. Define
\begin{align*}
H=&~\begin{pmatrix}
\frac{\partial g_1}{\partial x_1}&\cdots&\frac{\partial g_1}{\partial x_{n-1}}\\
\vdots &\ddots & \vdots\\
\frac{\partial g_{n-1}}{\partial x_1}&\cdots &\frac{\partial g_{n-1}}{\partial x_{n-1}}\\
\end{pmatrix},\\
b=&~\left(\frac{\partial g_1}{\partial x_n},...,\frac{\partial g_{n-1}}{\partial x_n}\right)^T,\\
c=&~\left(\frac{\partial}{\partial x_1}[g(x)\cdot x],...,\frac{\partial}{\partial x_{n-1}}[g(x)\cdot x]\right),\\
\hat{f}=&~(f_1,...,f_{n-1})^T,\ \hat{g}=(g_1,...,g_{n-1})^T.
\end{align*}
Because $g_n\equiv 1$, to differentiate $y=f(g(y),g(y)\cdot y)$ in $y$ at $y=x$, we have that\footnote{$I_n$ denotes the identity matrix.}
\[I_n=D_pf(p,m)Dg(x)+D_mf(p,m)D[g(x)\cdot x]=F\times \begin{pmatrix}
H & b\\
c & \frac{\partial}{\partial x_n}[g(x)\cdot x].
\end{pmatrix}\]
Therefore, $F$ is regular and
\begin{align*}
F^{-1}=&~\begin{pmatrix}
H & b\\
c & \frac{\partial}{\partial x_n}[g(x)\cdot x]
\end{pmatrix}\\
=&~\begin{pmatrix}
I_{n-1} & 0\\
\hat{f}^T & 1
\end{pmatrix}\times
\begin{pmatrix}
H & b\\
\hat{g}^T & 1
\end{pmatrix}\\
=&~\begin{pmatrix}
I_{n-1} & 0\\
\hat{f}^T & 1
\end{pmatrix}\times
\begin{pmatrix}
H-b\hat{g}^T & b\\
0 & 1
\end{pmatrix}\times
\begin{pmatrix}
I_{n-1} & 0\\
\hat{g}^T & 1
\end{pmatrix}.
\end{align*}
Because $H-b\hat{g}^T=A_g(x)$ by definition, we have that $A_g(x)$ is regular, and
\begin{align*}
\hat{S}=&~F\times \begin{pmatrix}
I_{n-1} & 0\\
\hat{f}^T & 1
\end{pmatrix}\\
=&~\begin{pmatrix}
I_{n-1} & 0\\
\hat{g}^T & 1
\end{pmatrix}^{-1}\times
\begin{pmatrix}
A_g(x) & b\\
0 & 1
\end{pmatrix}^{-1}\\
=&~\begin{pmatrix}
I_{n-1} & 0\\
-\hat{g}^T & 1
\end{pmatrix}\times
\begin{pmatrix}
(A_g(x))^{-1} & -(A_g(x))^{-1}b\\
0 & 1
\end{pmatrix}
\end{align*}
as desired. This completes the proof. $\blacksquare$

\vspace{12pt}
Now, let $g(x)$ be defined by (\ref{GALE2}) and choose any $x\in \tilde{\Omega}$. Let $k(x)=\frac{1}{g_3(x)}g(x)$, $p=k(x)$ and $m=k(x)\cdot x$. Then, $f^G(p,m)=x$. Clearly $f^G$ is differentiable at $(p,m)$, and $k$ is differentiable at $x$.

Because $f^G$ is homogeneous of degree zero, we have that
\[D_pf^G(p,m)p+D_mf^G(p,m)m=0.\]
On the other hand, because $f^G$ satisfies Walras' law,
\[m=(f^G)^T(p,m)p.\]
To combine these equations, we obtain
\[S_{f^G}(p,m)p=0.\]
By Lemma 6, we have that $\tilde{S}_{f^G}(p,m)$ is regular, and thus the rank of $S_{f^G}(p,m)$ is $2$.

Suppose that there exists $(q,w)\in \mathbb{R}^3_{++}\times \mathbb{R}_{++}$ such that $f^G(q,w)=x$ and $q$ is not proportional to $p$. Choose any $t\in [0,1]$, and define $(r,c)=(1-t)(p,m)+t(q,w)$. Suppose that $f^G(r,c)=y\neq x$. By Walras' law, $r\cdot x=c$. Because $r\cdot y\le c$, either $p\cdot y\le m$ or $q\cdot y\le w$, which contradicts the weak axiom. Therefore, we have that $f^G(r,c)=x$ for all $t\in [0,1]$. Hence,
\[S_{f^G}(p,m)(q-p)=\lim_{t\downarrow 0}\frac{f^G((1-t)p+tq,(1-t)m+tw)-f^G(p,m)}{t}=0.\]
This implies that the dimension of the kernel of $S_{f^G}(p,m)$ is greater than or equal to $2$, which contradicts the rank-nullity theorem. Therefore, such a $(q,w)$ is absent, and we conclude that $\tilde{P}=\{(mg(x),m)|x\in\tilde{\Omega},\ m>0\}$.

By the way, suppose that $f$ is a CoD that satisfies Walras' law, $g$ is an inverse demand function of $f$, $x=f(p,m)$, $f$ is differentiable at $(p,m)$, and $g$ is differentiable at $x$. Using Lemmas 3, 5, and 6, we can easily show that $g$ violates (\ref{JACOBI}) at $x$ if and only if $S_f(p,m)$ is not symmetric. Let $x=(1,1,1)$ and $(p,m)=(1,1,1,3)$. Because $g(x)$ defined by (\ref{GALE2}) violates (\ref{JACOBI}) at $x$, $S_{f^G}(p,m)$ is not symmetric, and by Theorem 3, we can conclude that $f^G$ is not a demand function. This is another proof that Gale's example is not a demand function.

However, this logic can be used because $f$ is differentiable. To prove III) of Theorem 4, we need a more complicated argument because $f^{\succsim^g}$ is not necessarily differentiable.

\subsection{Proof of I) of Theorem 4}

First, we must define several symbols. In this definition, we abbreviate variables, but all symbols are actually functions on $\tilde{\Omega}^2$. For any $(x,v)\in \tilde{\Omega}^2$, we define\footnote{We restrict the operator $R$ to $\mbox{span}\{x,v\}$.}
\[a_1=\frac{1}{\|x\|}x,\]
\[a_2=\begin{cases}
\frac{1}{\|v-(v\cdot a_1)a_1\|}(v-(v\cdot a_1)a_1) & \mbox{if }v\neq (v\cdot a_1)a_1,\\
0 & \mbox{otherwise},
\end{cases}\]
\[Py=(y\cdot a_1)a_1+(y\cdot a_2)a_2,\]
\[Rw=(w\cdot a_1)a_2-(w\cdot a_2)a_1,\]
\[v_1=\arg\min\{w\cdot a_1|w\in P\mathbb{R}^n_+,\ \|w\|=1,\ w\cdot a_2\ge 0\},\]
\[v_2=\arg\min\{w\cdot a_1|w\in P\mathbb{R}^n_+,\ \|w\|=1,\ w\cdot a_2\le 0\}.\]
\[\Delta=\{w\in \mbox{span}\{x,v\}|w\cdot Rv\le 0,\ w\cdot v_1\ge x\cdot v_1,\ w\cdot v_2\le x\cdot v_2\}.\]
\[C=\|x\|\|v-(v\cdot a_1)a_1\|.\]
Moreover, we define $y_1=y_2=x$ if $x$ is proportional to $v$, and otherwise, $y_i$ is the unique intersection of $\{s_1v|s_1\in \mathbb{R}\}\cap \{x+s_2Rv_i|s_2\in \mathbb{R}\}$.

\vspace{12pt}
\noindent
{\bf Lemma 7}. All the above symbols are well-defined. Moreover, the following results hold.
\begin{enumerate}[i)]
\item If $x$ is not proportional to $v$, then $\{a_1,a_2\}$ is the orthonormal basis of $\mbox{span}\{x,v\}$ derived from $x,v$ by the Gram-Schmidt method, and $P$ is the orthogonal projection from $\mathbb{R}^n$ onto $\mbox{span}\{x,v\}$.\footnote{As a result, $y\cdot w=Py\cdot w$ for any $y\in \mathbb{R}^n$ and $w\in \mbox{span}\{x,v\}$.}

\item If $x$ is not proportional to $v$, then $R$ is the unique orthogonal transformation on $\mbox{span}\{x,v\}$ such that $Ra_1=a_2$ and $Ra_2=-a_1$. Moreover, if $T$ is an orthogonal transformation on $\mbox{span}\{x,v\}$ such that $w\cdot Tw=0$ for any $w\in \mbox{span}\{x,v\}$, then we must have either $T=R$ or $T=-R=R^{-1}=R^3$.\footnote{In particular, if $z\in \mbox{span}\{x,v\}\cap \tilde{\Omega}$ and $[x,z]\cap \{cv|c\in \mathbb{R}\}=\emptyset$, then $R(x,v)=R(z,v)$. Note that $[x,z]$ represents $\{(1-t)x+tz|t\in [0,1]\}$.}

\item If $x$ is not proportional to $v$, then both $v_1$ and $v_2$ are continuous and single-valued at $(x,v)$. Moreover, $P\mathbb{R}^n_+=\{c_1v_1+c_2v_2|c_1,c_2\ge 0\}$.

\item Both $y_1$ and $y_2$ are continuous and single-valued. Moreover, for any $(x,v)\in \tilde{\Omega}^2$, $y_1(x,v),\ y_2(x,v)\in \tilde{\Omega}$.

\item $\Delta=(x+RP\mathbb{R}^n_+)\cap \{w\in \mbox{span}\{x,v\}|w\cdot Rv\le 0\}=\mbox{co}\{x,y_1,y_2\}$.\footnote{By this result, we have that $\Delta$ is a compact subset of $\tilde{\Omega}$.}

\item $(y\cdot x)v-(y\cdot v)x=C\cdot RPy$ for any $y\in \mathbb{R}^n$.
\end{enumerate}

\vspace{12pt}
We omit the proof of Lemma 7. See the proof of Theorem 1 in Hosoya (2013) for more detailed arguments.

We now prove I) of Theorem 4. For any $(x,v)\in \tilde{\Omega}^2$, let $y(\cdot;x,v)$ be the nonextendable solution to the following initial value problem:
\begin{equation}\label{INTEG}
\dot{y}(t)=(g(y(t))\cdot x)v-(g(y(t))\cdot v)x,\ y(0)=x.
\end{equation}
Note that, by Lemma 7, this equation can be rewritten as
\begin{equation}\label{INTEG2}
\dot{y}(t)=CRPg(y(t)),\ y(0)=x.
\end{equation}
Recall the following definition:
\[w^*=(v\cdot x)v-(v\cdot v)x=CRv,\]
\[t(x,v)=\inf\{t\ge 0|y(t;x,v)\cdot w^*\ge 0\}.\]
We first show the well-definedness of $t(x,v)$. Actually, $t(x,v)=0$ if $x$ is proportional to $v$, and thus we can assume that $x$ is not proportional to $v$. By the Cauchy-Schwarz inequality, we have that
\[x\cdot w^*<0.\]
Moreover,
\[\dot{y}(t;x,v)\cdot w^*=(CRPg(y(t;x,v)))\cdot (CRv)=C^2g(y(t;x,v))\cdot v>0.\]
Furthermore, because $\dot{y}(t;x,v)\in RP\mathbb{R}^n_+$, by v) of Lemma 7, we have that $y(t;x,v)\in \Delta$ if and only if $t\ge 0$ and $y(t;x,v)\cdot w^*\le 0$. Because $\Delta$ is compact and $y(t;x,v)$ is nonextendable, we have that there exists $t^*>0$ such that $y(t^*;x,v)\notin\Delta$, which implies that $y(t^*;x,v)\cdot w^*>0$. This indicates that $t(x,v)$ is well-defined. Moreover, clearly $y(t;x,v)\cdot w^*=0$ if and only if $t=t(x,v)$, which implies that $y(t;x,v)$ is proportional to $v$ if and only if $t=t(x,v)$. Because $t(x,v)$ is the unique solution to the following equation:
\[y(t;x,v)\cdot w^*=0,\]
by the implicit function theorem, the function $t:(y,z)\mapsto t(y,z)$ is $C^k$ around $(x,v)$, and thus $u^g$ is also $C^k$ around $(x,v)$. By iv) of Lemma 7, we can easily verify that $u^g$ is continuous on $\tilde{\Omega}^2$,\footnote{Recall that $u^g(x,v)v\in [y_1(x,v),y_2(x,v)]$ for any $(x,v)\in\tilde{\Omega}^2$.} and thus $\succsim^g$ is a continuous binary relation on $\tilde{\Omega}$.

It suffices to show that $\succsim^g$ is complete, p-transitive, and monotone. We introduce a lemma.

\vspace{12pt}
\noindent
{\bf Lemma 8}. Suppose that $x,v,z,w\in \Omega$, $x$ is not proportional to $v$, $z$ is not proportional to $w$, and $\mbox{span}\{x,v\}=\mbox{span}\{z,w\}$. If there exist $t_1,t_2\in \mathbb{R}$ such that $y(t_1;x,v)=y(t_2;z,w)$, then the trajectory of $y(t;x,v)$ is the same as that of $y(t;z,w)$.

\vspace{12pt}
\noindent
{\bf Proof}. Since $P(x,v)=P(z,w)$, we abbreviate both $P(x,v)$ and $P(z,w)$ as $P$. We abbreviate $R(x,v)$ as $R$. By ii), $R(z,w)$ is equal to either $R$ or $-R$. We define $s=1$ if $R(z,w)=R$ and $s=-1$ otherwise. Define
\[z_1(t)=y((C(x,v))^{-1}t;x,v), z_2(t)=y(s(C(z,w))^{-1}t;z,w).\]
For vi),
\[\dot{z}_1(t)=RPg(z_1(t)),\ \dot{z}_2(t)=RPg(z_2(t)).\]
Hence, both $z_1$ and $z_2$ are solutions to the following autonomous differential equation:
\[\dot{z}=RPg(z).\]
Moreover, we can easily check that both $z_1$ and $z_2$ are nonextendable. By assumption, there exist $t_3,t_4$ such that $z_1(t_3)=z_2(t_4)$, and thus the trajectory of $z_1$ must be equal to that of $z_2$. This completes the proof. $\blacksquare$

\vspace{12pt}
Now, we prove that 
\begin{equation}\label{KEY}
u^g(y,v)\ge u^g(z,v)\Leftrightarrow u^g(y,z)\ge 1(\Leftrightarrow y\succsim^gz)
\end{equation}
for all $x,v\in \tilde{\Omega}$ and $y,z\in\mbox{span}\{x,v\}\cap \tilde{\Omega}$.\footnote{That is, the function $y\mapsto u^g(y,v)$ represents $\succsim^g$ on $\tilde{\Omega}\cap \mbox{span}\{x,v\}$.} If $x$ is proportional to $v$, then (\ref{KEY}) holds trivially. Hence, we assume that $x$ is not proportional to $v$. If $z$ is proportional to $v$, then $z=av$, and by Lemma 8, $y(t(y,z);y,z)=y(t(y,v);y,v)$. Therefore, $u^g(z,v)=a$ and $u^g(y,v)=au^g(y,z)$, which implies that
\[u^g(y,v)\ge u^g(z,v)\Leftrightarrow au^g(y,z)\ge a\Leftrightarrow u^g(y,z)\ge 1,\]
as desired.

Hence, we assume that $z$ is not proportional to $v$. Suppose that $y$ is proportional to $z$. If $u^g(y,z)=1$, then $y=z$, and thus $u^g(y,v)=u^g(z,v)$. Suppose that $u^g(y,z)>1$. Define
\[c(s)=\frac{\|y(t(y,(1-s)y+sv);y,(1-s)y+sv)\|}{\|y(t(z,(1-s)z+sv);z,(1-s)z+sv)\|}.\]
By assumption, $c(0)=u^g(y,z)>1$. If $c(1)\le 1$, then $c(s)=1$ for some $s\in [0,1]$ by the intermediate value theorem. Thus, there exist $t',t''$ such that $y(t';y,v)=y(t'';z,v)$ and $c(s)\equiv 1$ by Lemma 8. This implies that $u^g(y,z)=c(0)=1$, which is a contradiction. Hence, $c(1)>1$, and thus, $u^g(y,v)>u^g(z,v)$. Finally, if $u^g(y,z)<1$, then by the symmetrical argument as above, we can show that $u^g(y,v)<u^g(z,v)$, and thus (\ref{KEY}) holds when $y$ is proportional to $z$.

Next, suppose that $y$ is not proportional to $z$. By Lemma 8, the trajectory of $y(t;y,v)$ is the same as that of $y(t;u^g(y,z)z,v)$, and thus
\begin{align}
u^g(u^g(y,z)z,v)=&~\frac{\|y(t(u^g(y,z)z,v);u^g(y,z)z,v)\|}{\|v\|}\nonumber \\
=&~\frac{\|y(t(y,v);y,v)\|}{\|v\|}\label{KEY2}\\
=&~u^g(y,v).\nonumber
\end{align}
Because we have already shown that $u^g(u^g(y,z)z,v)\ge u^g(z,v)$ if and only if $u^g(y,z)\ge 1$, we have that (\ref{KEY}) holds.\footnote{We mention that (\ref{KEY2}) holds unconditionally: that is, if $\dim(\mbox{span}\{y,z,v\})\le 2$, then
\[u^g(u^g(y,z)z,v)=u^g(y,v)\]
even when $y$ is proportional to $z$ or $z$ is proportional to $v$. The proof is easy and thus omitted.}

Completeness and p-transitivity of $\succsim^g$ directly follow from (\ref{KEY}).\footnote{For completeness, note that by (\ref{KEY}),
\[u^g(x,v)\ge 1\Leftrightarrow u^g(v,x)\le 1.\]}

It suffices to show that $\succsim^g$ is monotone. Actually, we prove that
\[v\gneq x\Rightarrow v\succ^gx.\]
Let $(x,v)\in \tilde{\Omega}^2$ and $v\gneq x$. If $x$ is proportional to $v$, then clearly $v\succ^gx$. Hence, we assume that $x$ is not proportional to $v$. By the completeness of $\succsim^g$, it suffices to show that $u^g(x,v)<1$. Define $z=v-x$. Then, $z\in \mbox{span}\{x,v\}$ and $z\gneq 0$. By i), ii), and vi) of Lemma 7,
\[\dot{y}(t;x,v)\cdot Rz>0,\]
and thus,
\begin{align*}
u^g(x,v)v\cdot Rz=&~y(t(x,v);x,v)\cdot Rz\\
>&~y(0;x,v)\cdot Rz\\
=&~x\cdot Rz\\
=&~v\cdot Rz.
\end{align*}
Therefore, to prove that $u^g(x,v)<1$, it suffices to show that $v\cdot Rz<0$.

Because $x\cdot a_2=0$,
\begin{align*}
v\cdot Rz=&~v\cdot [(z\cdot a_1)a_2-(z\cdot a_2)a_1]\\
=&~(z\cdot a_1)(v\cdot a_2)-(v\cdot a_2)(v\cdot a_1)\\
=&~-(v\cdot a_2)(x\cdot a_1).
\end{align*}
Since $x\cdot a_1>0$ by definition and $v\cdot a_2\ge 0$ by the Cauchy-Schwarz inequality, we have that $v\cdot Rz\le 0$. If $v\cdot Rz=0$, then $v$ is proportional to $z$, and thus, $x$ is proportional to $v$, which contradicts our initial assumption. Hence, we have that $v\cdot Rz<0$, which completes the proof of I). $\blacksquare$

\subsection{Proof of II) of Theorem 4}
Let $D_xu^g$ denote the partial derivative of $u^g$ with respect to the first variable or its transpose.\footnote{We think that this abbreviation of the transpose symbol does not cause any confusion.} We first present a lemma.

\vspace{12pt}
\noindent
{\bf Lemma 9}. Suppose that $(x,v)\in \tilde{\Omega}^2$ and $x$ is not proportional to $v$. Then, the following results hold.
\begin{enumerate}[1)]
\item $D_xu^g(x,v)\neq 0$ and $D_xu^g(x,v)x>0$.

\item $PD_xu^g(x,v)=\lambda(x)Pg(x)$, where $\lambda(x)=\frac{D_xu^g(x,v)x}{g(x)\cdot x}>0$.
\end{enumerate}

\vspace{12pt}
\noindent
{\bf Proof}. If $y,z,w\in \tilde{\Omega}$ are linearly dependent, then by (\ref{KEY2}), $u^g(y,w)=u^g(u^g(y,z)z,w)$. Therefore, for any $t>0$,
\[t=u^g(tv,v)=u^g(u^g(tv,x)x,v),\ 1=u^g(x,x)=u^g(u^g(x,v)v,x).\]
Taking derivatives with respect to $t$ on both sides of the former equation at $t=u^g(x,v)$, we have that,
\begin{align*}
1=&~D_xu^g(u^g(u^g(x,v)v,x)x,v)x\times [D_xu^g(u^g(x,v)v,x)v]\\
=&~D_xu^g(x,v)x\times [D_xu^g(u^g(x,v)v,x)v],
\end{align*}
which implies that $D_xu^g(x,v)\neq 0$ and $D_xu^g(x,v)x\neq 0$. Since $u^g(tx,v)>u^g(x,v)$ for any $t>1$, $D_xu^g(x,v)x>0$ and 1) holds.

To prove 2), define $z=\dot{y}(0;x,v)$. By (\ref{KEY2}), $u^g(y(t;x,v),v)=u^g(x,v)$. Taking derivatives with respect to $t$ at $t=0$, we have that $D_xu^g(x,v)z=0$. Note that because $z=CRPg(x)$, $g(x)\cdot z=0$, and thus both $Pg(x)$ and $PD_xu^g(x,v)$ belong to $\{w\in \mbox{span}\{x,v\}|w\cdot z=0\}$, whose dimension is exactly $1$. Hence, these are linearly dependent, and thus 2) holds. This completes the proof. $\blacksquare$

\vspace{12pt}
Suppose that $x\in f^{\succsim^g}(g(x),g(x)\cdot x)$ for all $x\in \tilde{\Omega}$. Choose any $x\in \tilde{\Omega}$ and $w\in \mathbb{R}^n$ such that $w\cdot g(x)=0$. If $w=0$, then clearly $w^TDg(x)w=0$. Otherwise, we can assume without loss of generality that $x+w,x-w\in \tilde{\Omega}$. Let $v=x+w$, $x(t)=x+tw$ and $c(t)=u^g(x(t),v)$. Since $x\in f^{\succsim^g}(g(x),g(x)\cdot x)$, we have that $c(0)\ge c(t)$ for any $t\in [-1,1]$. Note that $c'(t)=D_xu^g(x(t),v)w$ for all $t\in ]-1,1[$. By the mean value theorem, there exists a positive sequence $(t_m)$ such that $t_m\downarrow 0$ as $m\to \infty$ and $c'(t_m)\le 0$ for all $m$. Then,
\begin{align*}
0\ge&~\limsup_{m\to \infty}\frac{D_xu^g(x(t_m),v)w}{t_m}\\
=&~\limsup_{m\rightarrow \infty}\frac{\lambda(x(t_m))g(x(t_m))\cdot w}{t_m}\\
\ge&~M\limsup_{m\rightarrow\infty}\frac{g(x(t_m))\cdot w}{t_m}\\
=&~M\limsup_{m\rightarrow\infty}\frac{g(x(t_m))-g(x)}{t_m}\cdot w\\
=&~M[w^TDg(x)w],
\end{align*}
where $M=\max_{t\in [0,\frac{1}{2}]}\lambda(x(t))>0$. Hence, $w^TDg(x)w\le 0$ for all $x\in \tilde{\Omega}$ and $w\in \mathbb{R}^n$ such that $w\cdot g(x)=0$.

Second, suppose that there exists $x\in \tilde{\Omega}$ such that $x\notin f^{\succsim^g}(g(x),g(x)\cdot x)$. By p-transitivity, continuity, and monotonicity of $\succsim^g$, there exists $v\in\tilde{\Omega}$ such that $g(x)\cdot v<g(x)\cdot x$ and $v\sim^g x$. Clearly, $v$ is not proportional to $x$. Define $x(t)=(1-t)x+tv$. Then,
\begin{align*}
\left.\frac{d}{dt}[u^g(x(t),v)]\right|_{t=0}=&~D_xu^g(x,v)(v-x)\\
=&~\lambda(x)g(x)\cdot (v-x)\\
=&~\lambda(x)[g(x)\cdot v-g(x)\cdot x]<0,
\end{align*}
and thus, $u^g(x(t),v)<u^g(x,v)$ for a sufficiently small $t>0$, which implies that $x(t)\not\succsim^gx$, and thus $\succsim^g$ is not convex.

Hence, for proving the first claim of II) of Theorem 4, it suffices to show that if $w^TDg(x)w\le 0$ for all $x\in \tilde{\Omega}$ and $w\in \mathbb{R}^n$ such that $w\cdot g(x)=0$, then $\succsim^g$ is convex. Suppose not. Then, there exist $x,v\in \tilde{\Omega}$ and $t\in ]0,1[$ such that $v\sim^gx$ and $(1-t)x+tv\not\succsim^gx$. Clearly, $v$ is not proportional to $x$. Let $w=v-x$, $x(s)=x+sw$, and
\[s^*=\max[\arg\min\{u^g(x(s),v)|s\in [0,1]\}].\]
Since $u^g(x(t),v)<1$, we have $s^*\in ]0,1[$. Define $c(s)=u^g(x(s),v)$. Then, $c(s^*)\le c(s)$ for any $s\in [0,1]$, and thus, $c'(s^*)=0$. Hence, $D_xu^g(x(s^*),v)w=0$. Let $p=PD_xu^g(x(s^*),v)$. Then, $p\cdot w=0$. By 1) of Lemma 9, $p\cdot x(s^*)>0$, and thus $p\neq 0$.

Consider the following equation:
\[f(a,b)=u^g(bp+x(s^*+a),v)=u^g(x(s^*),v).\]
Note that $\frac{\partial f}{\partial b}(0,0)=\|p\|^2\neq 0$. Applying the implicit function theorem, we can prove the existence of the $C^k$ function $b:]-\varepsilon,\varepsilon[\to \mathbb{R}$ such that $b(0)=0$ and $f(a,b(a))=u^g(x(s^*),v)$ for any $a\in ]-\varepsilon,\varepsilon[$. Differentiating both sides with respect to $a$, we have
\[D_xu^g(b(a)p+x(s^*+a),v)[b'(a)p+w]=0.\]
Hence,
\[0=D_xu^g(x(s^*),v)[b'(0)p+w]=b'(0)\|p\|^2,\]
which implies that $b'(0)=0$. By Lemma 9, if $a>0$ is so small that $D_xu^g(b(a)p+x(s^*+a),v)p>0$, then
\begin{align*}
0=&~\limsup_{a'\downarrow a}\frac{1}{a'-a}[-D_xu^g(b(a)p+x(s^*+a),v)[b'(a')p+w]\\
&~+D_xu^g(b(a)p+x(s^*+a),v)[b'(a')p+w]]\\
=&~\limsup_{a'\downarrow a}\frac{1}{a'-a}[-\lambda(b(a)p+x(s^*+a))g(b(a)p+x(s^*+a))\cdot[b'(a')p+w]\\
&~+D_xu^g(b(a)p+x(s^*+a),v)[b'(a')p+w]]\\
=&~\limsup_{a'\downarrow a}\frac{1}{a'-a}[\lambda(b(a)p+x(s^*+a))\\
&~\times [g(b(a')+x(s^*+a'))-g(b(a)p+x(s^*+a))]\cdot[b'(a')p+w]\\
&~+D_xu^g(b(a)p+x(s^*+a),v)[(b'(a')p+w)-(b'(a)p+w)]]\\
=&~\lambda(b(a)p+x(s^*+a))[b'(a)p+w]^TDg(b(a)p+x(s^*+a))[b'(a)p+w]\\
&~+D_xu^g(b(a)p+x(s^*+a),v)p\times \limsup_{a'\downarrow a}\frac{b'(a')-b'(a)}{a'-a},
\end{align*}
and thus, $\limsup_{a'\downarrow a}\frac{b'(a')-b'(a)}{a'-a}\ge 0$. By the same arguments, we can show that $\limsup_{a'\uparrow a}\frac{b'(a')-b'(a)}{a'-a}\ge 0$.

Now, choose any such $a>0$. Define $h(\theta)=b'(\theta)a-b'(a)\theta$. Then, $h(0)=h(a)=0$, and thus, there exists $\theta^*\in ]0,a[$ such that either $h(\theta^*)\ge h(\theta)$ for any $\theta\in [0,a]$ or $h(\theta^*)\le h(\theta)$ for any $\theta\in [0,a]$. If the former holds,
\begin{align*}
0\ge&~\limsup_{\theta\downarrow \theta^*}\frac{h(\theta)-h(\theta^*)}{\theta-\theta^*}\\
=&~a\limsup_{\theta\downarrow \theta^*}\frac{b'(\theta)-b'(\theta^*)}{\theta-\theta^*}-b'(a),
\end{align*}
and thus we have that $b'(a)\ge 0$. By the symmetrical arguments, we can show that $b'(a)\ge 0$ in the latter case. Hence, we have that $b'(a)\ge 0$ for a sufficiently small $a>0$, and thus, $b(a)\ge 0$ for a sufficiently small $a>0$.

On the other hand, $D_xu^g(x(s^*),v)p=\|p\|^2>0$, and thus,
\[D_xu^g(b'p+x(s^*+a),v)p>0\mbox{ for any }b'\in [0,b(a)],\]
for a sufficiently small $a>0$. Therefore,
\[u^g(x(s^*+a),v)\le u^g(b(a)p+x(s^*+a),v)=u^g(x(s^*),v),\]
which contradicts the definition of $s^*$. Hence, the first claim of II) is correct.

Next, suppose that $w^TDg(x)w<0$ for all $x\in \mathbb{R}^n_{++}$ and $w\in \mathbb{R}^n$ such that $w\neq 0$ and $w\cdot g(x)=0$. Choose any $(p,m)\in \mathbb{R}^n_{++}\times\mathbb{R}_{++}$. For any $x\in \tilde{\Omega}$ such that $p\cdot x=m$, if $g(x)$ is not proportional to $p$, then there exists $v\in \tilde{\Omega}$ such that $p\cdot v\le p\cdot x$ and $v\succ^gx$, and thus, $x\notin f^{\succsim^g}(p,m)$. Therefore, if there is no $x$ such that $p\cdot x=m$ and $g(x)$ is proportional to $p$, then $f^{\succsim^g}(p,m)=\emptyset$. Next, suppose that $p=ag(x)$ for some $a>0$ and $m=p\cdot x$. We have already shown that $x\in f^{\succsim^g}(p,m)$. Suppose that $\succsim^g$ is strictly convex. Then, $x\succ^gv$ for any $v\in \tilde{\Omega}$ such that $p\cdot v\le m$ and $v\neq x$, which implies that $f^{\succsim^g}(p,m)=\{x\}$. Hence, it suffices to show that $\succsim^g$ is strictly convex. Suppose not. Then, there exist $y,z\in\tilde{\Omega}$ and $t\in ]0,1[$ such that $y\sim^gz$, $y\neq z$ and $(1-t)y+tz\not\succ^gy$. Define $y(s)=(1-s)y+sz$. We have already shown that $\succsim^g$ is convex, and thus $y(s)\succsim^gy$ for any $s\in [0,1]$. This implies that $u^g(y(t),y)\le u^g(y(s),y)$ for any $s\in [0,1]$. By the first-order condition and Lemma 9,
\[g(y(t))\cdot (z-y)=0,\]
and by the mean value theorem, there exists a sequence $(s_m)$ such that $s_m\downarrow t$ as $m\to \infty$ and
\[g(y(s_m))\cdot (z-y)\ge 0\]
for all $m$. Threfore, 
\[(z-y)^TDg(y(t))(z-y)=\lim_{m\to \infty}\frac{[g(y(s_m))-g(y(t))]\cdot (z-y)}{s_m-t}\ge 0,\]
which contradicts our initial assumption. This completes the proof of II). $\blacksquare$

\subsection{Proof of III) of Theorem 4}
We separate the proof into two steps.

\vspace{12pt}
\noindent
{\bf Step 1}. Suppose that $g:\tilde{\Omega}\to \mathbb{R}^n_{++}$ is $C^k$. Then, ii)-v) of III) of Theorem 4 are equivalent.

\vspace{12pt}
\noindent
{\bf Proof of Step 1}. First, we show that iii) is equivalent to iv). Suppose that iii) holds. Then, for any $x,z\in\tilde{\Omega}$,
\[\frac{d}{dt}[u^g_v(y(t;x,z))]=0,\]
and thus $u^g_v(x)=u^g_v(u^g(x,z)z)$. Hence,
\[u^g_v(x)\ge u^g_v(z)\Leftrightarrow u^g_v(u^g(x,z)z)\ge u^g_v(z)\Leftrightarrow u^g(x,z)\ge 1\Leftrightarrow x\succsim^gz,\]
which implies that iv) holds.

Conversely, suppose that iv) holds. Choose any $x\in \tilde{\Omega}$ and any linearly independent family $v_1,...,v_{n-1}\in\mathbb{R}^n$ such that $v_i\cdot g(x)=0$ and $x+v_i\in\tilde{\Omega}$ for all $i\in \{1,...,n-1\}$. Since both $v_i$ and $\dot{y}(0;x,x+v_i)$ are orthogonal to $P(x,x+v_i)g(x)$, $\dot{y}(0;x,x+v_i)$ is proportional to $v_i$. Since $y(t;x,x+v_i)\sim^gx$ for any $t$, we have that $\nabla u^g_v(x)$ is orthogonal to $\dot{y}(0;x,x+v_i)$, and thus $\nabla u^g_v(x)\cdot v_i=0$ for all $i\in \{1,...,n-1\}$. Therefore, $\nabla u^g_v(x)=\lambda(x)g(x)$ for some $\lambda(x)\in\mathbb{R}$, which implies that iii) holds.

Second, we show that iv) is equivalent to ii). It is obvious that iv) implies ii). Conversely, suppose that ii) holds. First, choose any $x,z,v\in \tilde{\Omega}$. Then,
\[u^g(x,v)v\sim^gx\sim^gu^g(x,z)z\sim^gu^g(u^g(x,z)z,v)v,\]
which implies that $u^g(x,v)=u^g(u^g(x,z)z,v)$.

Next, choose any $x,z,v\in \tilde{\Omega}$ and suppose that $u_v^g(x)\ge u_v^g(z)$. Then,
\[u^g(u^g(x,z)z,v)=u^g(x,v)\ge u^g(z,v),\]
which implies that $u^g(x,z)\ge 1$. Hence, $x\succsim^gz$. Conversely, suppose that $x\succsim^gz$. Then, $u^g(x,z)\ge 1$, and thus,
\[u_v^g(x)=u^g(u^g(x,z)z,v)\ge u_v^g(z).\]
Hence, $u_v^g$ represents $\succsim^g$.

Fix any $v\in \tilde{\Omega}$ and choose any $z\in \tilde{\Omega}$ such that $z$ is not proportional to $v$. Then, for any $x\in \tilde{\Omega}$, we have $x$ is not proportional to either $v$ or $z$. If $x$ is not proportional to $v$, clearly $u^g_v$ is $C^k$ around $x$. Otherwise,
\[u_v^g(y)=u^g(u^g(y,z)z,v)\]
for any $y\in \tilde{\Omega}$ and the right-hand side is $C^k$ in $y$ around $x$. Therefore, $u_v^g$ is $C^k$ on $\tilde{\Omega}$, and thus iv) holds.

It suffices to show that iii) is equivalent to v). By Frobenius' theorem, iii) implies v). Conversely, suppose that v) holds. We introduce a lemma.

\vspace{12pt}
\noindent
{\bf Lemma 10}. For any $v\in \tilde{\Omega}$, there exist open neighborhoods $W^v,W_1^v$ of $v$, $\hat{u}^v:W_1^v\to \mathbb{R}$, and $\lambda^v:W_1^v\to \mathbb{R}_{++}$ such that

\begin{enumerate}[1)]
\item $u^g(x,z)=u^g(u^g(x,y)y,z)$ for any $x,y,z\in W^v$,

\item $y([0,t(x,y)];x,y)\subset W_1^v$ for any $x,y\in W^v$,

\item $\hat{u}^v$ is $C^k$ on $W_1^v$ and $\nabla\hat{u}^v(x)=\lambda^v(x)g(x)$ for any $x\in W_1^v$.
\end{enumerate}

\vspace{12pt}
\noindent
{\bf Proof}. Fix any $v\in \tilde{\Omega}$. By Frobenius' theorem, there exists an open and convex neighborhood $W_1^v\subset \tilde{\Omega}$ of $v$, $\hat{u}^v:W_1^v\to \mathbb{R}$, and $\lambda^v:W_1^v\to \mathbb{R}_{++}$ such that $\hat{u}^v$ is $C^k$ and $\nabla\hat{u}^v(x)=\lambda^v(x)g(x)$ for all $x\in W_1^v$. Since $y_1,y_2$ are continuous on $\tilde{\Omega}^2$ and $y_1(v,v)=v=y_2(v,v)$, $U_1^v=y_1^{-1}(W_1^v)\cap y_2^{-1}(W_1^v)$ is open and includes $(v,v)$.\footnote{See Lemma 7.} Let $W_2^v\subset W_1^v$ be an open, convex neighborhood of $v$ such that $W_2^v\times W_2^v\subset U_1^v$, $U_2^v=y_1^{-1}(W_2^v)\cap y_2^{-1}(W_2^v)$, and $W^v\subset W_2^v$ be an open neighborhood of $v$ such that $W^v\times W^v\subset U_2^v$.

Clearly, 3) holds. To show 2), choose any $x,y\in W^v$. If $x$ is proportional to $y$, then $t(x,y)=0$, and thus 2) holds. Otherwise, since $x,y\in W^v$, we have that $(x,y)\in U_2^v$, and thus, $y_1(x,y),y_2(x,y)\in W_2^v$. Since $W_2^v$ is convex, we have that $\Delta(x,y)\subset W_2^v\subset W_1^v$, and thus, $y([0,t(x,y)];x,y)\subset \Delta(x,y)\subset W_1^v$. Hence, 2) holds.

To show 1), suppose that $x,y,z\in W^v$. If these vectors are linearly dependent, then we have shown that $u^g(x,z)=u^g(u^g(x,y)y,z)$ holds in (\ref{KEY2}). Otherwise, we have already shown that $y([0,t(x,y)];x,y)\subset W_2^v$, and thus, $u^g(x,y)y\in W_2^v$. Hence, $y_1(u^g(x,y)y,z),y_2(u^g(x,y)y,z)\in W_1^v$, and thus, $\Delta(u^g(x,y)y,z)\subset W_1^v$. Therefore, $y([0,t(u^g(x,y)y,z)];u^g(x,y)y,z)\subset W_1^v$. Moreover, we have already proved that $y([0,t(x,z)];x,z)\subset W_1^v$.

By easy computations,
\[\frac{d}{dt}\hat{u}^v(y(t;x,y))=0\mbox{ for any }t\in [0,t(x,y)],\]
\[\frac{d}{dt}\hat{u}^v(y(t;u^g(x,y)y,z))=0\mbox{ for any }t\in [0,t(u^g(x,y)y,z)],\]
and
\[\frac{d}{dt}\hat{u}^v(y(t;x,z))=0\mbox{ for any }t\in [0,t(x,z)].\]
Hence, we obtain
\[\hat{u}^v(u^g(x,z)z)=\hat{u}^v(x)=\hat{u}^v(u^g(x,y)y)=\hat{u}^v(u^g(u^g(x,y)y,z)z).\]
Since $\nabla\hat{u}^v(w)\gg 0$ for all $w\in W_1^v$, we have that $u^g(x,z)=u^g(u^g(x,y)y,z)$. Hence, 1) holds. This completes the proof. $\blacksquare$

\vspace{12pt}
Choose any $(x,v)\in \tilde{\Omega}^2$. If $x$ is not proportional to $v$, then $u^g$ is $C^k$ around $(x,v)$. Otherwise, there exists $s>0$ such that $x=sv$. Let $W^x$ be the set defined in Lemma 10, and choose any $z\in W^x$ that is not proportional to $x$. Let $W'=W^x\setminus \{tz|t\in\mathbb{R}\}$ and $W''=s^{-1}W'$. Then, $W'$ is an open neighborhood of $x$, $W''$ is an open neighborhood of $v$, and for any $(y,w)\in W'\times W''$, 
\[u^g(y,w)=su^g(y,sw)=su^g(u^g(y,z)z,sw)=u^g(u^g(y,z)z,w),\]
where the right-hand side is $C^k$ around $(x,v)$. Hence, $u^g$ is $C^k$ on $\tilde{\Omega}^2$.

Define
\[X=\{x\in \tilde{\Omega}|\exists \mu(x)\in\mathbb{R}\mbox{ such that }D_xu^g(x,v)=\mu(x)g(x)\}.\]
To prove iii), it suffices to show that $X=\tilde{\Omega}$.

First, we show that $X$ includes some open neighborhood of $L=\{sv|s>0\}$. Fix any $s>0$. Let $x=sv$ and choose $W^x, W_1^x, \hat{u}^x, \lambda^x$ as in Lemma 10. Fix any $y\in W^x$. Choose any $z\in W^x$ such that $\hat{u}^x(z)=\hat{u}^x(y)$. Since
\[\frac{d}{dt}\hat{u}^x(y(t;y,z))\equiv 0\]
for all $t\in [0,t(y,z)]$, we have that $\hat{u}^x(y)=\hat{u}^x(u^g(y,z)z)$, and thus, $u^g(y,z)=1$. Hence,
\[u^g(y,v)=su^g(y,x)=su^g(u^g(y,z)z,x)=su^g(z,x)=u^g(z,v).\]
Conversely, choose any $z\in W^x$ such that $u^g(y,v)=u^g(z,v)$. Then,
\[u^g(z,v)=u^g(y,v)=su^g(y,x)=su^g(u^g(y,z)z,x)=u^g(u^g(y,z)z,v),\]
which implies that $u^g(y,z)=1$. Hence,
\[\hat{u}^x(y)=\hat{u}^x(u^g(y,z)z)=\hat{u}^x(z).\]
Therefore, $\hat{u}^x(z)=\hat{u}^x(y)$ if and only if $u^g(z,v)=u^g(y,v)$.

By the preimage theorem, the set $(\hat{u}^x)^{-1}(y)$ is an $n-1$ dimensional $C^k$ manifold and both $\nabla\hat{u}^x(y)$ and $D_xu^g(y,v)$ are in the orthogonal complement of the tangent space $T_y((\hat{u}^x)^{-1}(y))$.\footnote{See Section 1.4 of Guillemin and Pollack (1974).} Hence, there exists $c\in \mathbb{R}$ such that
\[D_xu^g(y,v)=c\nabla\hat{u}^x(y)=c\lambda^x(y)g(y).\]
Therefore, we have proved that $\cup_{s>0}W^{sv}\subset X$.

Second, fix any $x\in \tilde{\Omega}$. It suffices to show that $x\in X$. If $x$ is proportional to $v$, then it is clear. Otherwise, define $A_x$ as the trajectory of $y(\cdot;x,v)$ and $B_x=A_x\cap Y$, where $Y$ is the interior of $X$. It suffices to show that $x\in B_x$. To show this, we prove that $B_x=A_x$. Since $A_x$ is connected, it suffices to show that $B_x$ is nonempty, open, and closed in $A_x$. Clearly, $B_x$ is open in $A_x$. Since $u^g(x,v)v\in B_x$, $B_x$ is nonempty. Hence, it suffices to show that $B_x$ is closed in $A_x$.

Suppose that $(z^m)$ is a sequence on $B_x$ that converges to $z^*\in A_x$. It suffices to show that $z^*\in B_x$. Without loss of generality, we assume that $z^*\neq u^g(x,v)v$. Choose $W^{z^*},W_1^{z^*},\hat{u}^{z^*},\lambda^{z^*}$ as in Lemma 10.

Since $W^{z^*}$ is open in $\tilde{\Omega}$, there exists $m$ such that $z^m\in W^{z^*}$. By the definition of $A_x$ and Lemma 8, there exists $t^m$ such that $y(t^m;z^*,v)=z^m$. Since $z^m\in B_x$, there exists an open set $W_{z^m}\subset W^{z^*}$ and $\mu:W_{z^m}\to\mathbb{R}$ such that $D_xu^g(z,v)=\mu(z)g(z)$ for any $z\in W_{z^m}$. Let $W_3$ be an open neighborhood of $z^*$ such that $W_3\subset W^{z^*}$ and $y(t^m;z,v)\in W_{z^m}$ for any $z\in W_3$.

By the local submersion theorem,\footnote{Again, see Section 1.4 of Guillemin and Pollack (1974).} there exist open $W_4,V\subset \mathbb{R}^n$, and $\phi:V\rightarrow W_4$ such that $z^*\in W_4\subset W_3$, $V$ is convex, $\phi$ is a bijection from $V$ onto $W_4$, both $\phi$ and $\phi^{-1}$ are $C^k$, and $(\hat{u}^{z^*}\circ \phi)(w)=w^1$ for any $w\in V$.

Now, suppose that $y,z\in W_4$, and $\hat{u}^{z^*}(y)=\hat{u}^{z^*}(z)$. Define $z_1(s)$ and $z_2(s)$ such that
\[z_1(s)=\phi((1-s)\phi^{-1}(y)+s\phi^{-1}(z)),\ z_2(s)=y(t^m;z_1(s),v).\]
Then, $\hat{u}^{z^*}(z_2(s))=\hat{u}^{z^*}(z_1(s))=\hat{u}^{z^*}(y)$ for any $s\in [0,1]$. On the other hand, $z_2(s)\in W_{z^m}$ for any $s\in [0,1]$, and thus,
\[\frac{d}{ds}[u^g(z_2(s),v)]=0,\]
which implies that
\[u^g(y,v)=u^g(z_1(0),v)=u^g(z_2(0),v)=u^g(z_2(1),v)=u^g(z_1(1),v)=u^g(z,v).\]
Symmetrically, we can show that if $u^g(y,v)=u^g(z,v)$, then $\hat{u}^{z^*}(y)=\hat{u}^{z^*}(z)$. Hence, $u^g(y,v)=u^g(z,v)$ if and only if $\hat{u}^{z^*}(y)=\hat{u}^{z^*}(z)$, and thus, we can prove $W_4\subset X$ using the preimage theorem. Therefore, v) implies iii), as desired. This completes the proof of Step 1. $\blacksquare$

\vspace{12pt}
\noindent
{\bf Step 2}. Suppose that $f:P\to \tilde{\Omega}$ is a CoD such that $g$ is an inverse demand function of $f$, and $f=f^{\succsim^g}$. Then, i) and ii) in III) of Theorem 4 are equivalent.

\vspace{12pt}
\noindent
{\bf Proof of Step 2}. Clearly ii) implies i). Conversely, suppose that i) holds. Then, there exists a weak order $\succsim$ such that $f^{\succsim^g}=f^{\succsim}$. First, we note the following fact. Choose any $x,v\in \tilde{\Omega}$, and define 
\[x_0^m=x,\]
\[x_{i+1}^m=x_i^m+\frac{t(x,v)}{m}CRPg(x_i^m).\]
The finite sequence $x_0^m,...,x_m^m$ is the explicit Euler approximation of (\ref{INTEG}), and thus $x_m^m\to y(t(x,v);x,v)$ as $m\to \infty$. Moreover, because $g(x_i^m)\cdot x_i^m=g(x_i^m)\cdot x_{i+1}^m$, we have that $x_i^m\succ x_{i+1}^m$, and thus $x\succ x_m^m$. This implies that if $0<c<u^g(x,v)$, then $x\succ cv$.

Now, suppose that ii) is violated. Then, there exist $x,y,z\in \tilde{\Omega}$ such that $x\succsim^gy$, $y\succsim^gz$, but $z\succ^gx$. This implies that $u^g(z,x)>1$. Choose any $a\in ]1,u^g(z,x)[$. Then, $z\succ ax$. Because $ax\gg x$, we have that $u^g(ax,y)>1$, and thus there exists $b\in ]1,u^g(ax,y)[$. Then, $ax\succ by$, and thus $z\succ by$. Because $by\gg y$, we have that $u^g(by,z)>1$, and thus $by\succ z$, which is a contradiction. This completes the proof of Step 2. $\blacksquare$

\vspace{12pt}
Clearly, Steps 1-2 imply III) of Theorem 4. This completes the proof. $\blacksquare$

\subsection{Proof of Theorem 5}
Suppose that $\succsim^g$ is transitive. By Step 1 in the proof of III) of Theorem 4, it is represented by $u^g_v$, and $u^g_v$ is $C^1$. By Lagrange's multiplier rule, we have that
\[\nabla u^g_v(x)=\lambda(x)g(x)\]
for some $\lambda(x)>0$. Now, choose any piecewise $C^1$ closed curve $x:[0,T]\to \tilde{\Omega}$. Then,
\[u^g_v(x(0))=u^g_v(x(T)),\]
and thus, there exists a non-null set $S\subset [0,1]$ such that for all $t\in S$,
\[\frac{d}{dt}u^g_v(x(t))\le 0.\]
This implies that for all $t\in S$,
\[g(x(t))\dot{x}(t)\le 0,\]
and thus $g$ satisfies Ville's axiom.

Conversely, suppose that $\succsim^g$ is not transitive. Then, there exists $x,y,z\in \tilde{\Omega}$ such that $x\succsim^gy$, $y\succsim^gz$, and $z\succ^gx$. Because $\succsim^g$ is p-transitive, we have that $x,y,z$ are linearly independent. Because $z\succ^gx$, we have that $u^g(x,z)<1$. Consider the following differential equation
\[\dot{z}^1(t;a)=(g(z^1(t;a))\cdot x)z-(g(z^1(t;a))\cdot z)x+az,\ z^1(0;a)=x.\]
We have that $z^1(t;0)=y(t;x,z)$, and thus for sufficiently small $a>0$, there exists $t_1>0$ such that $z^1(t_1;a)=\alpha z$ for some $\alpha\in ]0,1[$. Note that,
\[g(z^1(t;a))\cdot \dot{z}^1(t;a)=az\cdot g(z^1(t;a))>0\]
for all $t\in [0,t_1]$. Because $u^g(y,z)\ge 1$, by (\ref{KEY}), we have that $u^g(z,y)\le 1$, and thus $u^g(\alpha z,y)<1$. Consider the following differential equation
\[\dot{z}^2(t;b)=(g(z^2(t;b))\cdot \alpha z)y-(g(z^2(t;b))\cdot y)\alpha z+by,\ z^2(0;b)=\alpha z.\]
We have that $z^2(t;0)=y(t;\alpha z,y)$, and thus for sufficiently small $b>0$, there exists $t_2>0$ such that $z^2(t_2;b)=\beta y$ for some $\beta\in ]0,1[$. Again note that,
\[g(z^2(t;b))\cdot \dot{z}^2(t;b)=by\cdot g(z^2(t;b))>0\]
for all $t\in [0,t_2]$. Because $u^g(x,y)\ge 1$, we have that $u^g(y,x)\le 1$, and thus $u^g(\beta y,x)<1$. Consider the following differential equation
\[\dot{z}^3(t;c)=(g(z^3(t;c))\cdot \beta y)x-(g(z^3(t;c))\cdot x)\beta y+cx,\ z^3(0;c)=\beta y.\]
We have that $z^3(t;0)=y(t;\beta y,x)$, and thus for sufficiently small $c>0$, there exists $t_3>0$ such that $z^3(t_3;c)=\gamma x$ for some $\gamma\in ]0,1[$. Note that
\[g(z^3(t;c))\cdot \dot{z}^3(t;c)=cx\cdot g(z^3(t;c))>0\]
for all $t\in [0,t_3]$. Finally, define $z^4(t)=[(1-t)\gamma+t]x$, and
\[x(t)=\begin{cases}
z^1(t;a) & \mbox{if }0\le t\le t_1,\\
z^2(t-t_1;b) & \mbox{if }t_1\le t\le t_1+t_2,\\
z^3(t-t_1-t_2;c) & \mbox{if }t_1+t_2\le t\le t_1+t_2+t_3,\\
z^4(t-t_1-t_2-t_3) & \mbox{if }t_1+t_2+t_3\le t\le t_1+t_2+t_3+1.
\end{cases}\]
Then, for $T=t_1+t_2+t_3+1$, $x:[0,T]\to \tilde{\Omega}$ is a piecewise $C^1$ closed curve, and for almost all $t\in [0,T]$,
\[g(x(t))\cdot \dot{x}(t)>0,\]
which implies that $g$ violates Ville's axiom. This completes the proof. $\blacksquare$

\section*{Reference}

\begin{description}
\item{[1]} Chambers, C. P. and Echenique, F. (2016) \textit{Revealed Preference Theory}. Cambridge University Press, Cambridge.

\item{[2]} Debreu, G. (1952) ``Definite and Semi-Definite Quadratic Forms.'' Econometrica 20, 295-300.

\item{[3]} Debreu, G. (1954) ``Representation of a Preference Ordering by a Numerical Function.'' In: Thrall, R. M., Coombs, C. H., Davis, R. L. (Eds.) \textit{Decision Processes.} Wiley, New York, 159-165.

\item{[4]} Dieudonne, J. (1969) \textit{Foundations of Modern Analysis}. Academic Press, London.

\item{[5]} Gale, D. (1960) ``A Note on Revealed Preference.'' Economica 27, 348-354.

\item{[6]} Guillemin, V. and Pollack, A. (1974) \textit{Differential Topology}. Prentice Hall, New Jersey.

\item{[7]} Hosoya, Y. (2013) ``Measuring Utility from Demand.'' Journal of Mathematical Economics 49, 82-96.

\item{[8]} Hosoya, Y. (2019) ``Revealed Preference Theory.'' Applied Analysis and Optimization 3, 179-204.

\item{[9]} Hosoya, Y. (2020) ``Recoverability Revisited.'' Journal of Mathematical Economics 90, 31-41.

\item{[10]} Hosoya, Y. (2021a) ``Equivalence between Nikliborc's Theorem and Frobenius' Theorem.'' Pure and Applied Functional Analysis 6, 719-741.

\item{[11]} Hosoya, Y. (2021b) ``The Weak Axiom of Revealed Preference and Inverse Problems in Consumer Theory.'' Linear and Nonlinear Analysis 7, 9-31.

\item{[12]} Houthakker, H. S. (1950) ``Revealed Preference and the Utility Function.'' Economica 17, 159-174.

\item{[13]} Hurwicz, L. and Richter, M. K. (1979a) ``Ville Axioms and Consumer Theory.'' Econometrica 47, 603-619.

\item{[14]} Hurwicz, L. and Richter, M. K. (1979b) ``An Integrability Condition with Applications to Utility Theory and Thermodynamics.'' Journal of Mathematical Economics 6, 7-14.

\item{[15]} Pareto, V. (1906a) \textit{Manuale di Economia Politica con una Introduzione alla Scienza Sociale}. Societa Editrice Libraria, Milano.

\item{[16]} Pareto, V. (1906b) ``L'ofelimit\`a Nei Cicli Non Chuisi.'' Giornale degli Economisti 33, 15-30. English translated by Chipman, J. S. ``Ophelimity in Nonclosed Cycle.'' in: Chipman, J. S., Hurwicz, L., Richter, M. K., Sonnenschein, H. F. (Eds.) \textit{Preferences, Utility, and Demand}. Harcourt Brace Jovanovich, New York, pp.370-385 (1971).

\item{[17]} Pareto, V. (1909) \textit{Manuel d'Economie Politique}. Giard et E. Briere, Paris.

\item{[18]} Richter, M. K. (1966) ``Revealed Preference Theory.'' Econometrica 34, 635-645.

\item{[19]} Rose, H. (1958) ``Consistency of Preference: The Two-Commodity Case.'' Review of Economic Studies 25, 124-125.

\item{[20]} Samuelson, P. A. (1938) ``A Note on the Pure Theory of Consumer's Behaviour''. Economica 5, 61-71.

\item{[21]} Samuelson, P. A. (1950) ``The Problem of Integrability in Utility Theory.'' Economica 17, 355-385.

\item{[22]} Suda, S. (2007) ``Vilfredo Pareto and the Integrability Problem of Demand Function.'' Keio Journal of Economics 99, 637-655.

\item{[23]} Szpilrajn, E. (1930) ``Sur l'Extension de l'Ordre Partiel.'' Fundamenta Mathematicae 16, 386-389.

\item{[24]} Uzawa, H. (1959) ``Preferences and Rational Choice in the Theory of Consumption.'' In: Arrow, K. J., Karlin, S., Suppes, P. (Eds.) \textit{Mathematical Models in the Social Science, 1959: Proceedings of the First Stanford Symposium}, Stanford University Press, Stanford, pp.129-149. Reprinted in: Chipman, J. S., Hurwicz, L., Richter, M. K., Sonnenschein, H. F. (Eds.) \textit{Preferences, Utility, and Demand}. Harcourt Brace Jovanovich, New York, pp.7-28 (1971).

\item{[25]} Volterra, V. (1906) ``L'economia Matematica ed il Nuovo Manuale del Prof. Pareto.'' Giornale degli Economisti 33, 296-301. English translated by Kirman, A. P. ``Mathematical Economics and Professor Pareto's New Manual.'' in: Chipman, J. S., Hurwicz, L., Richter, M. K., Sonnenschein, H. F. (Eds.) \textit{Preferences, Utility, and Demand}. Harcourt Brace Jovanovich, New York, pp.365-369 (1971).
\end{description}

\end{document}